\documentstyle[psfig]{aa}
\def\zabs{$z_{\rm abs}$}

\def\mgii{Mg~{\sc ii}~}

\def\mgiia{Mg~{\sc ii}$\lambda$2796~ }

\begin{document}
\thesaurus{11.17.1;11.17.4 APM08279+5255}
\title{Structure of the Mg~{\sc ii} and damped Lyman-$\alpha$ systems 
along the line of sight to APM~08279+5255
\thanks{Based on observations collected at the W.M. Keck Observatory, 
which is operated as a
scientific partnership among the California Institute of Technology, the
University of California and the National Aeronautics and Space
Administration. The Observatory was made possible by the generous financial
support of the W.M. Keck Foundation.}
}
\author{Patrick Petitjean\inst{1,2} \and Bastien Aracil\inst{1} 
\and R. Srianand$^3$ \and Rodrigo Ibata$^4$}
\institute{$^1$Institut d'Astrophysique de Paris -- CNRS, 98bis Boulevard 
Arago, F-75014 Paris, France\\
$^2$UA CNRS 173 -- DAEC, Observatoire de Paris-Meudon, F-92195 Meudon
Cedex, France \\
$^3$IUCAA, Post Bag 4, Ganesh Khind, Pune 411 007, India \\
$^4$Max-Plank Institut f\"ur Astronomie, K\"onigstuhl 17, D-69117 Heidelberg,
Germany}
\date{ }
\offprints{Patrick Petitjean}
\maketitle
\markboth{}{}
\begin{abstract}
A study of the absorption  systems toward the gravitationally lensed quasar
APM~08279+5255  is presented. \par\noindent
Most of  the  Mg~{\sc ii}
systems  in the redshift range $z$~$\sim$~1.2--2.07,  
although saturated, show large residuals at the bottom of the lines.
The most likely interpretation is 
that individual clouds within Mg~{\sc ii} 
halos do cover only one of the two  brightest QSO images. The separation 
between the  two lines  of  sight decreases from  1.7  to  
0.7~$h^{-1}_{75}$~kpc ($q_{\rm  o}$~=~0.5, $z_{\rm lens}$~=~1) between 
$z$~=~1.22 and $z$~=~2.07. This reveals 
that Mg~{\sc ii} halos are made of a collection of clouds of
radius smaller than about 1~$h^{-1}_{75}$~kpc. 
\par\noindent
Two  strong Mg~{\sc  ii} absorbers  at $z_{\rm  abs}$~=~1.062 and  1.181 are
studied in detail. This is the first
time that the Na~{\sc i}$\lambda$3303 doublet is detected 
in such high redshift 
systems. Together with the detection of the
Mg~{\sc i}$\lambda$2852 transition, this strongly constrains the physical
characteristics of the gas. The  $N$(Na~{\sc i})/$N$(Mg~{\sc i})
ratio is found to  be larger than  unity, implying that the  gas is cool and
neutral. The Doppler parameters measured in  individual and well
detached  components is probably as small as 1~km~s$^{-1}$.
The column densities of Na~{\sc i},  Ca~{\sc ii}, Mg~{\sc  i}, Ti~{\sc ii},
Mn~{\sc ii} and Fe~{\sc ii} observed at $z_{\rm abs}$~=~1.1801 are very close
to  that  observed along  the  line of  sight towards  23~Ori in  our
Galaxy.  The  shape of the QSO  continuum is consistent  with attenuation by
dust at $z$~$\sim$~1 ($A_{\rm V}$~$\sim$~0.5~mag).
Altogether it is found that the  H~{\sc i}  column  density at
$z$~=~1 is  of the  order of 1  to 5~10$^{21}$~cm$^{-2}$,  the corresponding
metallicity is in the range 1--0.3~$Z_{\odot}$, the overall dust-to-metal ratio
is about half that in our Galaxy and the relative depletion of
iron, titanium, manganese and calcium is
similar to what is observed in cool gas in the disk of our Galaxy.
The objects associated with these two systems could both contribute to the 
lens together with another possible strong system at $z_{\rm abs}$~=~1.1727
and the strong Lyman-$\alpha$ system at $z_{\rm abs}$~=~2.974.
\par\noindent
The probable damped Lyman-$\alpha$ system at $z_{\rm abs}$~=~2.974 has
19.8~$<$~log~$N$(H~{\sc  i})~$<$~20.3. The  transverse  dimension of  the
absorber is  larger than 200~$h^{-1}_{75}$~pc.  Column  densities of Al~{\sc
ii}, Fe~{\sc ii}, Si~{\sc ii},  C~{\sc ii} and O~{\sc i} indicate abundances
relative to  solar of  $-$2.31, $-$2.26, $-$2.10,  $-$2.35 and  $-$2.37 for,
respectively,    Fe,    Al,    Si,    C   and    O    (for    log~$N$(H~{\sc
i})~=~20.3).  These surprizingly similar values indicate that
the amount of dust in the cloud is very small as are any deviations from 
relative solar abundances. It seems likely that the  
upper limits found for the zinc metallicity of 
several damped Lyman-$\alpha$ systems at $z$~$>$~3 in previous surveys is 
indicative of a true cosmological evolution of  the metallicity in individual
systems.


\end{abstract}

\section {Introduction}
The  gravitationally lensed Broad  Absorption Line  (BAL) QSO
APM~08279+5255 ($z_{\rm em}$~=~3.911) has been given much attention since its 
discovery by Irwin et
al.  (1998), as it is one of  the most luminous objects in the universe even
after  correction  for  the  gravitational  lensing  induced  amplification.
Adaptive-optics  imaging has  revealed two  main components  (Ledoux  et al.
1998b), separated  by 0.378$\pm$0.001~arcsec  as measured on  HST/NICMOS data
(Ibata  et  al.   1999),   and  of  relative  brightness  $F_{\rm  B}/F_{\rm
A}$~=~0.773$\pm$0.007.  The HST  images reveal also the presence  of a third
object C  with $F_{\rm C}/F_{\rm A}$~=~0.175$\pm$0.008,  located in between A
and B and almost aligned with them.  The point-spread-function model fits on
the  three   objects  are  consistent   with  the  three   components  being
point-sources, and their colors are similar within the uncertainties.  There
is no trace of the lensing object up to magnitude $V$~=~23.

A high S/N ratio high-resolution  spectrum of APM 08279+5255 was obtained at
the  Keck telescope (Ellison  et al.   1999a,b), and  made available  to the
astronomical community. This spectrum, though complicated by the combination
of light traveling along three  different sightlines, is a unique laboratory
for studying the intervening and associated absorption systems.

In  this paper we  study the structure of six intervening Mg~{\sc ii}
systems at 1.2~$<$~$z$~$<$~2.07 and the physical characteristics of the gas 
in two very  strong  Mg~{\sc ii}  systems
detected  at $z_{\rm  abs}$~=~1.06 and  1.18,  which, we  argue, are  damped
Lyman-$\alpha$ systems and  may well reveal the lensing galaxies. We
also comment on a third  probable damped Lyman-$\alpha$  system at  
$z_{\rm abs}$~=~2.974.
This paper is organized as follows: the data are described in Section~2; 
the structure of the intervening Mg~{\sc ii} systems is investigated 
using the covering factor analysis in Section 3;
we demonstrate that  the Mg~{\sc ii}  systems at
$z_{\rm  abs}$~=~1.06   and  1.18  are  damped   Lyman-$\alpha$  systems  in
Section~4; we discuss a probable damped Lyman-$\alpha$  system  
at $z_{\rm abs}$~=~2.974 
in Section~5. We adopt $H_{\rm o}$~=~75~km~s$^{-1}$~Mpc$^{-1}$ and 
$q_{\rm  o}$~=~0.5 througout the paper. 

\section {Data}
A high S/N ratio,  high-resolution spectrum of the $z_{\rm em}$~=~3.911 
quasar APM 08279+5255  was obtained
with the HIRES echelle spectrograph  at the 10m Keck-I telescope (Ellison et
al.  1999a,b).   This data  was made public  together with  a low-resolution
spectrum of  the quasar and a  high-resolution spectrum of  a standard star.
We have  corrected the high-resolution spectrum of  APM~08279+5255 for small
discontinuities   in  the  continuum,   which  are   probably  due   to  the
inappropriate merging  of different orders. These  discontinuities have been
recognized by comparing the high  and low-resolution spectra.  The latter is
also  used  for  normalization  of the  high-resolution  data.   Atmospheric
absorption features were identified from the standard star spectrum.
Voigt profile fitting of the absorption features have been performed using
the context FITLYMAN (Fontana \& Ballester 1995) of the European Southern 
Observatory data reduction package MIDAS and the code VPFIT 
(Carswell et al. 1987).
We  have  measured the  final  spectral  resolution  by fitting  the  narrow
atmospheric  absorption   lines  which  are   free  of  blending.   We  find
$FWHM$~$\sim$~8~km~s$^{-1}$ ($b$~$\sim$~4.8~km~s$^{-1}$) at 6900~\AA,
$R$~=~37500, and use this value throughout the paper.

\section {Structure of the intervening $z$~$>$~1 Mg~{\sc ii} absorbers}
\begin{figure}
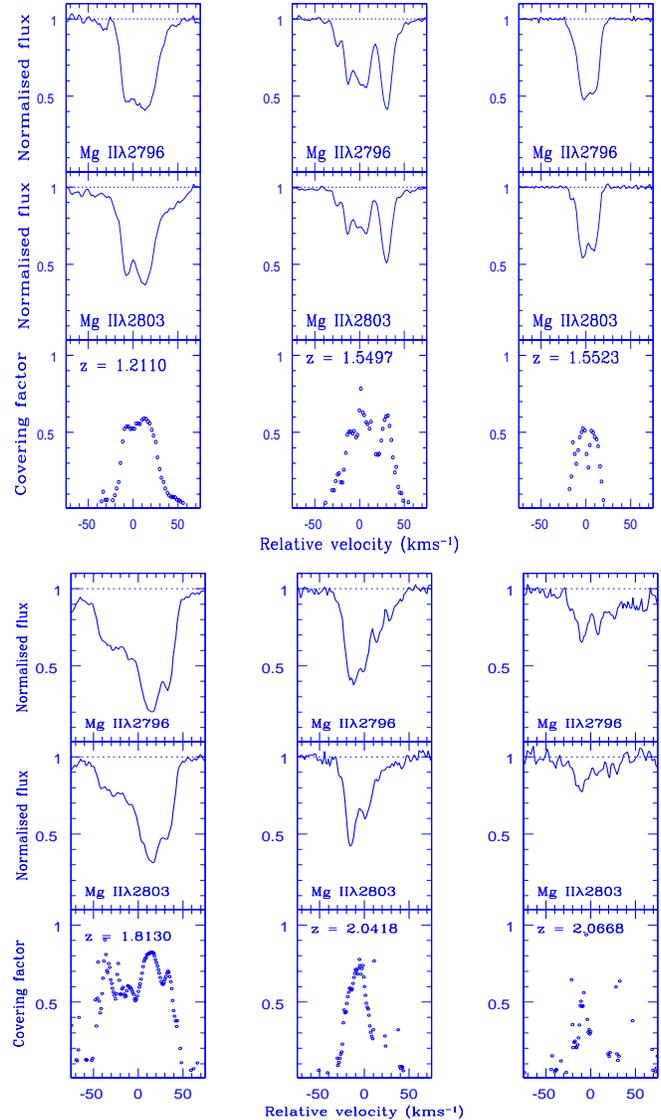

\centerline{\vbox{
\psfig{figure=9595.f1a,height=8.cm,width=9.3cm,angle=270}
\vskip -0.5cm
\psfig{figure=9595.f1b,height=8.cm,width=9.3cm,angle=0}
}}
\caption[]{Covering factor (bottom panels) of six Mg~{\sc ii} systems observed
along the line of sight to APM~08279+5255 calculated from the profiles
of the Mg~{\sc ii}$\lambda$2796 (top panels) and Mg~{\sc ii}$\lambda$2803
(middle panels) absorptions. }
\label{cf}
\end{figure}
Ellison et al. (1999b) have already noted  that the two lines of some
of the Mg~{\sc ii}$\lambda\lambda$2796,2803  doublets cannot  be fitted
with the same column density and Doppler parameter.   As can be seen
on Fig.~\ref{cf}, most of the systems have a doublet ratio close to
unity inspite of having residual intensities in the normalized spectrum
close to 0.5. 

If the background source is a point source and 
if the Mg~{\sc ii}$\lambda$2796 absorption line is resolved,
then the residual intensity of the normalized spectrum
measured at any velocity $v$ with respect
to the centroid of the line, is equal to $e^{-\tau(v)}$, where
$\tau(v)$ is the optical depth at $v$ and the residual intensity
of the Mg~{\sc ii}$\lambda$2803 line is $e^{-\tau(v)/2}$.
%
It is apparent from Fig.~\ref{cf} that the above condition is not
fullfilled for most of the Mg~{\sc ii} doublets. 

There are two possibilities to explain this.
If the lines are not resolved, the measured residual intensity is affected by  
convolution of the true absorption profile with the instrumental profile.
This can introduce artificial residuals at the bottom of the
saturated absorption features (see e.g. Lespine \& Petitjean 1997).  
Alternatively, as the observed light is a combination of light from
different images, it is possible that
the column densities are different along different lines of sight 
and, as a limiting case, that the absorbing cloud does not
cover all the images.
We discuss these different possibilities below.
\subsection {Unresolved narrow-components}
\begin{figure}
\centerline{\vbox{
\psfig{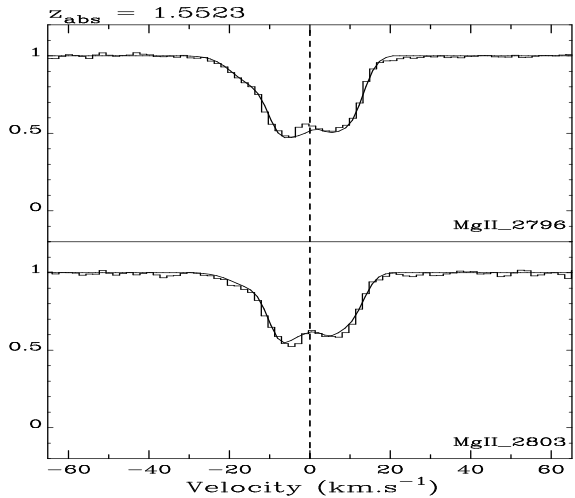}
}}
\caption[]{Best fit of the Mg~{\sc ii} doublet at $z_{\rm abs}$~=~1.5523
with five narrow components ($b$~$<$~1.5~km~s$^{-1}$) and a covering factor 
$f$~=~1 (reduced $\chi^2$~ of 1.5).}
\label{fit155a}
\end{figure}
\begin{figure}
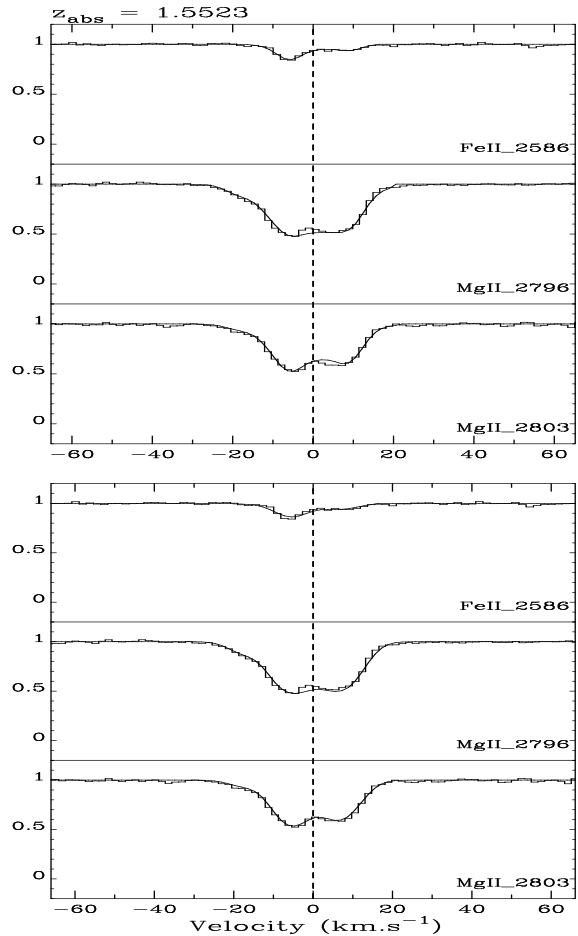

\centerline{\vbox{
\psfig{figure=9595.f3a,height=6.cm,width=7.5cm,angle=0}
\vskip 0.3cm
\psfig{figure=9595.f3b,height=6.cm,width=7.5cm,angle=0}
}}
\caption[]{Best fits of the $z_{\rm abs}$~=~1.5523 Mg~{\sc ii} and Fe~{\sc ii}
absorption lines obtained with five narrow ($b$~$<$~1.5~km~s$^{-1}$)
components and $f$~=~0.6 ({\sl top panel}, reduced $\chi^2$~ of 
0.9) and three broader ($b$~$>$~2.5~km~s$^{-1}$)
components and $f$~=~0.45 ({\sl bottom panel}, reduced $\chi^2$~ of 
0.6). The fit parameters are given in Table~1.}
\label{fit155b}
\end{figure}
\begin{figure}
\centerline{\vbox{
\psfig{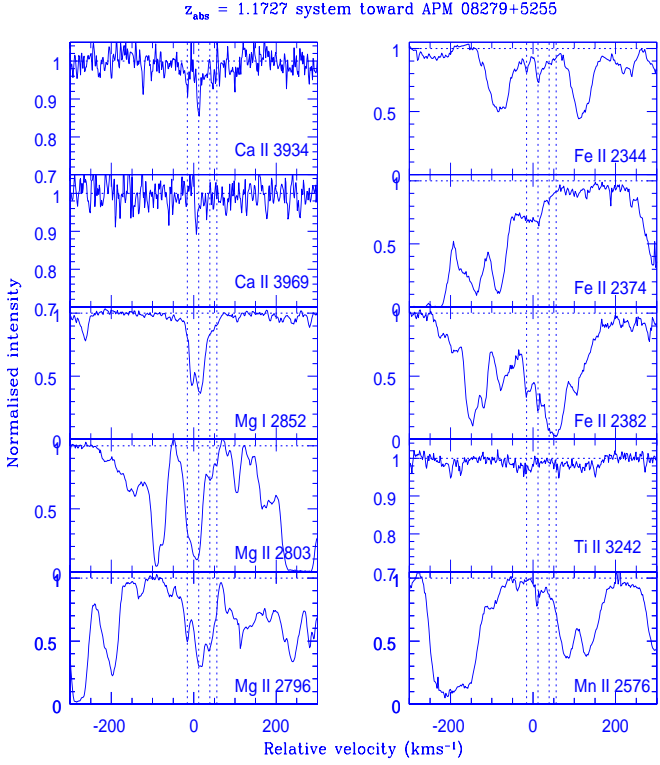}
}}
\caption[]{Probable Mg~{\sc ii} system at $z_{\rm abs}$~=~1.1727.
Most of the lines are blended but the similarity of the Fe~{\sc ii}, 
Mg~{\sc ii} and Ca~{\sc ii} profiles supports the identification.
Vertical dotted lines indicate the positions of the four probable components.
}
\label{s117}
\end{figure}
%
If the absorption profiles are made up of unresolved components,
the observed residual
intensity does not correspond to the real optical depth. 
Given the resolution of the spectrum ($R$~$\sim$~37500), a saturated 
Mg~{\sc ii}$\lambda$2796 line can have a residual intensity in the
normalized spectrum of 0.5 if its Doppler parameter is smaller than 
1.5~km~s$^{-1}$. In that case, the residual intensity of the 
Mg~{\sc ii}$\lambda$2803 line is in the range 0.5--0.6, depending on the 
actual column density. The two equivalent widths differ by no more than 20\%
(see e.g. Lespine \& Petitjean 1997). 

The system at $z_{\rm abs}$~=~1.5497 can be indeed fitted this way using 
10 components 
with $b$ values in the range 1.1--1.7~km~s$^{-1}$.
In that case the well detached cloud in the red wing (see Fig.~\ref{cf}) 
is fitted with two adhoc nearly identical components though the 
Mg~{\sc ii}$\lambda$2796 profile is 
perfectly fitted with a single {\sl resolved} component model. However
the one-component model cannot explain the strengths of the two
Mg~{\sc ii} lines without invoking partial coverage (see below).

We use the $z_{\rm abs}$~=~1.5523 system to illustrate the case. 
The final spectrum has a defect in the center of the 
Mg~{\sc ii}$\lambda$2796 line. We have therefore used 
three individual exposures of high S/N ratio
(Ellison private communication) to
correct for this. The final optical depth variations from one spectrum
to the other is about 2\%. 
Fig.~\ref{fit155a} shows the best fit to the doublet considering full
coverage ($f$~=~1). Five narrow components ($b$~$<$~1.5~km~s$^{-1}$)
are needed. Although good (reduced $\chi^2$ of 1.5), the 
fit is not completely satisfactory. 
We have fitted consistently the Mg~{\sc ii} together with the Fe~{\sc ii} lines
considering that only one of the brightest sources is covered. 
With five components, a good fit (reduced $\chi^2$ of 0.9),   
shown on Fig.~\ref{fit155b}, is obtained if B is not covered
($f$~=~0.6). 
Details of the subcomponent parameters are given in Table~1. As the 
Mg~{\sc ii} doublet ratio is close to one, small $b$ values are needed even 
with the assumption of partial coverage. 

It should be noted however that a Doppler parameter smaller than 
1.5~km~s$^{-1}$ corresponds to a temperature smaller than 3500~K, 
a surprizingly small temperature for this gas which is most probably 
ionized. Allowing for a smaller number of components, we can find 
a good fit with only three components, $f$~=~0.45 and 
reduced $\chi^2$~=~0.6 (see Fig.~\ref{fit155b} and Table~1). 
The $b$ values are larger than 2.5~km~s$^{-1}$ relaxing the restriction 
on the temperature. 

A statistically acceptable fit is difficult to find for
the systems at $z_{\rm abs}$~=~1.221 and 1.5523.
The red wing of the Mg~{\sc ii}$\lambda$2803 line at $z_{\rm abs}$~=~1.221
is blended with another absorption feature which we found difficult to
identify. It could be 
Mg~{\sc i}$\lambda$2852 at $z_{\rm abs}$~=~1.1727. This system is 
possibly detected by Fe~{\sc ii}$\lambda\lambda$2344,2382, 
Mn~{\sc ii}$\lambda$2576 and Ca~{\sc ii}$\lambda$3934
(see Fig.~\ref{s117}). The other lines are 
either below the detection limit or blended. 
Confirmation of the Ca~{\sc ii} lines, which are free of blending, would be 
particularly important as, probably, the presence of this additional system,
together with the presence of the three damped systems
at $z_{\rm abs}$~=~1.062, 1.181 and 2.974 (see below) 
should be taken into account in any model of the lens.

\begin{table}
\begin{tabular}{lllllll}
\multicolumn{7}{l}{{\bf Table 1.} Fit parameters for the
$z_{\rm abs}$~=~1.5523 system}\\ 
\hline
\multicolumn{1}{c}{$z$}&
\multicolumn{1}{c}{$b^b$}&\multicolumn{1}{c}{$\pm$}&
\multicolumn{1}{c}{$N$(MgII)$^a$}&
\multicolumn{1}{c}{$\pm$}&
\multicolumn{1}{c}{$N$(FeII)$^a$}&
\multicolumn{1}{c}{$\pm$}\\
\hline
\multicolumn{7}{c}{Best fit with five components; $f$~=~0.6}\\
1.552165 & 1.7 & 0.5 & 11.51 & 0.15 &       &     \\
1.552219 & 1.1 & 0.5 & 11.88 & 0.05 & 11.67 & 0.12 \\
1.552270 & 1.3 & 0.2 & 14.47 & 0.27 & 12.79 & 0.05 \\
1.552338 & 3.4 & 0.2 & 12.47 & 0.10 & 12.23 & 0.06 \\
1.552399 & 0.8 & 0.1 & 14.67 & 0.09 & 12.34 & 0.07 \\
\hline
\multicolumn{7}{c}{Best fit with three components; $f$~=~0.45}\\
1.552170 & 2.9 & 0.5 & 11.67 & 0.10 & 11.29 & 0.11 \\
1.552274 & 4.2 & 0.6 & 13.02 & 0.18 & 12.82 & 0.09 \\
1.552378 & 4.6 & 0.7 & 12.85 & 0.20 & 12.44 & 0.07 \\
\hline 
\multicolumn{7}{l}{$^a$ logarithm of the column density in cm$^{-2}$}\\
\multicolumn{7}{l}{$^b$ in km~s$^{-1}$}\\
\label{tab155}
\end{tabular}
\end{table}
\subsection{Partial covering factor}
\subsubsection{Computing a covering factor}
We can interpret the observations in terms  of a covering factor 
which is the fraction of the background source  covered by the absorbing 
cloud. The relative brightness of the three sources are 
$F_{\rm A, B, C}$/$F_{\rm tot}$~=~0.513, 0.397 and 0.090 (Ibata et al.  1999).
If one line-of-sight is completely  absorbed (condition imposed by the
fact that the doublets are saturated)
and the other free  of  absorption,  then the  covering  factor  is  
0.40, 0.49, 0.51 and 0.60 if, respectively, B only, B+C, A only and A+C are
covered (the case C only is very unlikely as C is close to A and
located in between A and B, Ibata et al. 1999; see however Srianand \&
Petitjean 2000). Given  the uncertainties, 
if  only  one  line-of-sight  is  completely  absorbed,  
the covering factor should be in the range 0.4--0.6.  Of course, 
it is larger if the second line-of-sight is not completely clear.
We have computed  the covering factor for the Mg~{\sc  ii} systems using the
method described by Srianand \&  Shankaranarayanan (1999). 
This assumes that the lines are resolved. It can be seen in
Fig.~\ref{cf}  that for the  three systems  with $z_{\rm  abs}$~$<$~1.7, the
covering  factor ranges between  0.5 and  0.6 whereas  for the  systems with
$z$~$>$~1.7, the  covering factor is  larger (but always less than 0.8)
with the possible  exception of the  $z_{\rm abs}$~=~2.0668  system. 
The  latter  system is  quite  weak however  and
uncertainties are  large.  The values of  the covering factor  for the three
lower redshift systems  suggest that the clouds cover one of the two
brightest background sources only. This has to be investigated in more
detail however.
\subsubsection{Optical depths along different sightlines}
\begin{figure*}
\centerline{\vbox{
\psfig{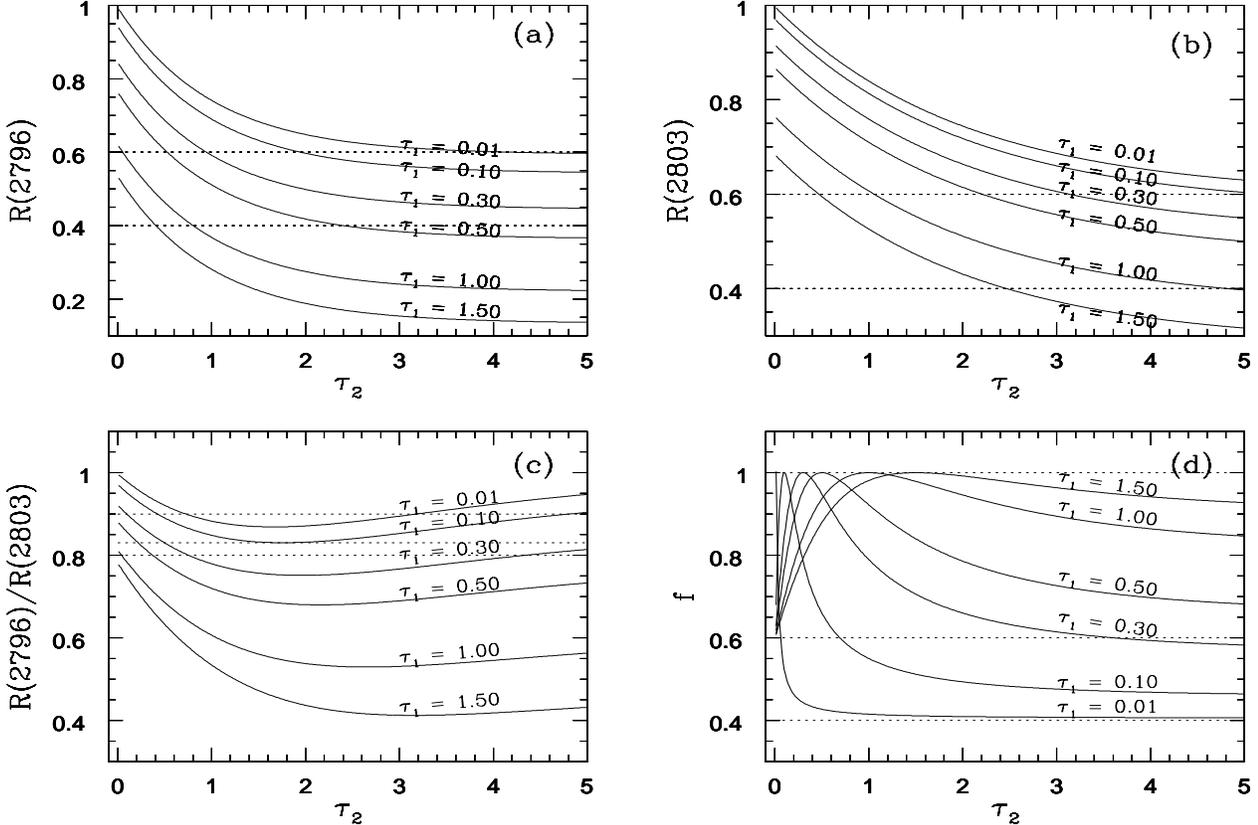}
}}
\caption[]{Panels (a) and (b) give, respectively, the observed residuals
$R$(2796) and $R$(2803) in the normalized spectrum and in the center of the 
Mg~{\sc ii}$\lambda$$\lambda$2796,2803 lines 
as a function of $\tau_2$ for different values of $\tau_1$. 
$\tau_1$ and $\tau_2$ are the optical depths in the center of the
Mg~{\sc ii}$\lambda$2796 line along, respectively, line of sight number one 
towards image A and line of sight number two towards image B
(see Eqs.~1 and 2). It is assumed that the fractional flux contributions
are $f_1$~=~0.6 and $f_2$~=~0.4.
The two horizontal dotted lines show the range of observed values.
Panel (c) gives the ratio of the residual intensities of the
two absorption lines as a function of $\tau_2$ for different values of 
$\tau_1$. The dotted lines show the observed values in the three systems 
at \zabs$<1.7$. In panel (d) the derived covering factor $f$ (Eq.~3)
is shown as a function of $\tau_1$ for different values of $\tau_2$. 
Measured values are indicated by dotted horizontal lines.}
\label{figg}
\end{figure*}
In this Section we investigate the effect of the optical depth 
being different along different sight lines.  In order
to make our analysis simpler, we consider only two images, A and B,
with fractional flux contributions $F_1$~=~0.6  and $F_2$~=~0.4. 
Suppose the optical depth along the two sight lines are $\tau_1$ 
and $\tau_2$ then the measured residual intensities are,
\begin{eqnarray}
R(2796) ~&=&~F_1 e^{-\tau_1} +F_2 e^{-\tau_2} \\ \nonumber 
R(2803) ~&=&~F_1 e^{-\tau_1/2} +F_2 e^{-\tau_2/2} 
\end{eqnarray}
%
The residual
intensities can be written as (Srianand \& Shankaranarayanan 1999),
\begin{eqnarray}
R(2796)~&=&~1-f +f e^{-\tau}\\ \nonumber
R(2803)~&=&~1-f +f e^{-\tau/2}
\end{eqnarray}
where, $f$  and $\tau$ are the resulting covering factor
and optical depth and therefore,
\begin{equation}
f~=~{[1-R(2803)]^2 \over 1+R(2796)-2R(2803) }
\end{equation}
From Eqs. (1) and (3) one can derive $f$ as a function of $F_1$,
$F_2$, $\tau_1$ and $\tau_2$. This analysis assumes that 
the absorption profiles are resolved in the HIRES spectrum. As 
$F_1$ and $F_2$ are known from observation, 
the covering factor depends on $\tau_1$ and $\tau_2$ only. 
 
In order to investigate the parameter space we have computed 
and plotted on Fig~\ref{figg}, the covering
factor (panel d), residual intensities  of the lines (panels a and b)
and their  ratio (panel c)  as a
function of  $\tau_2$  for different values of $\tau_1$. 
In panels (a) and (b), the two horizontal dotted lines show the range of 
observed values for the \mgii systems with \zabs$<$1.7. In  panel (c) 
the dotted lines give the measured values for the three low-redshift
Mg~{\sc ii} systems.  

As expected, when $\tau_1=\tau_2$, $f=1$ (albeit with various
ratios of residual intensities) and for $\tau_1\neq\tau_2$ the covering
factors are less than 1.0.  When $\tau_1\ge\tau_2$
(respectively $\tau_1\le\tau_2$), the covering factor is in the range
0.6--1.0 (respectively 0.4--1.0). Conversely,
the observed residual intensities,
$R$(2796) and $R$(2803), together with the measured covering factor, $f$, 
can be used to constrain $\tau_1$ and $\tau_2$. 

The covering factor estimates for \zabs$<$ 1.7 systems are in
the range 0.5 and 0.6 (see Fig.~\ref{cf}). This, together with the observed
residual intensities, indicates that the absorbing gas is saturated along
one sight line only with optical depth ratios as large as ten.
For example,
the well detached component in the red wing of the
\zabs~=~1.5497 system has $f~=~0.6$ (with a typical error 
of 0.02), a residual intensity ratio $\sim$0.84 (with a typical 
error of 0.02) and R(2796)$\simeq$0.40 at the core of the line.
This implies that the contribution to this absorption
comes mainly from the line of sight toward A+C with the
optical depth along B being more than an order of magnitude smaller.

\subsection {Equivalent width ratio}
The precise determination of the strength of the absorption 
lines along different lines of sight should await HST/STIS spectroscopic 
observations. Though we derived some information on this
in the previous Section using the absorption line residual intensities, 
the results depend crucially on the assumption that the absorption lines 
are resolved. 
We can complement the previous analysis using the total equivalent widths 
and their ratios without making any assumption about the spectral resolution 
In addition, when considering total equivalent widths, the consequence of 
contamination by weak lines is small
unlike in the case of the analysis of the residual intensities.
The constrains are much weaker however. 

\begin{figure}
\centerline{\vbox{
\psfig{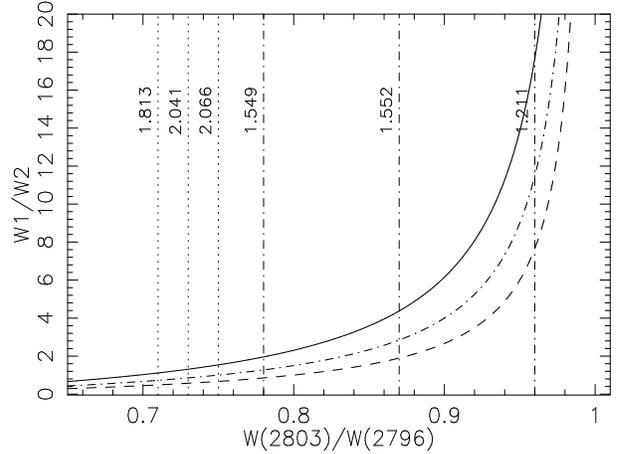}
}}
\caption[]{Equivalent width ratio of the Mg~{\sc ii}$\lambda$2796 
absorption line along two different line-of-sights 1 and 2 
versus the observed doublet ratio for three
different flux ratios $F_1$/$F_2$~=~0.65 (solid line), 1 (dashed-dotted
line) and 1.5 (dashed line). The Mg~{\sc ii} doublet is assumed to be
saturated along line-of-sight 1 and optically thin along line-of-sight 2.
The doublet ratios observed for six Mg~{\sc ii} systems toward 
APM~08279+5255 are shown as vertical
dashed-dotted and dotted lines for $z_{\rm abs}$~$<$~1.5 and 
$z_{\rm abs}$~$>$~1.5 respectively. Redshifts are indicated next to the
lines.}
\label{covering}
\end{figure}

Let us assume that the absorption is saturated along line of sight number one
and optically thin along line of sight number two.
Therefore $W_1^{\rm real}$(2796)~=~$W^{\rm real}_1$(2803) and $W_2^{\rm
real}$(2796)~=~2$\times$$W_2^{\rm real}$(2803). The combined observed
equivalent width ratio is:
\begin{equation}
{W^{\rm obs}(2803)\over W^{\rm obs}(2796)} = {
{1\over 2} + {W_1^{\rm real}(2796)\over W_2^{\rm real}(2796)}{F_1 \over F_2}
\over
1 + {W_1^{\rm real}(2796)\over W_2^{\rm real}(2796)}{F_1 \over F_2}
}
\end{equation}
where $F_1$ and $F_2$ are the fractional flux contributions of the two distinct 
background sources. In Fig.~\ref{covering}, we have plotted 
$W_1^{\rm real}$(2796) /$W_2^{\rm real}$(2796) versus
$W^{\rm obs}$(2803) /$W^{\rm obs}$(2796) for $F_1$/$F_2$ = 0.65
(1 is B; 2 is A+C), 1 (1 is A or B+C; 2 is B+C or A)
and 1.5 (1 is A+C; 2 is B). The vertical dashed-dotted lines correspond to the
observed doublet ratios of the systems at $z_{\rm abs}$~=~1.221, 1.5523
and of the reddest component of the $z_{\rm abs}$~=~1.5497 system.

From Fig.~\ref{covering}, it can be seen that the \zabs~=~1.221 system
(doublet ratio of 0.96) requires the ratio of the equivalent
widths of \mgiia along the two lines of sight to be larger than 8. and 
the column densities along the two sightlines differ by more 
than an order of magnitude.




%
\subsection {Dimension of individual clouds}
Rauch  et al.  (1999)  have observed  strong  variations of  C~{\sc
ii}  and Si~{\sc  ii} absorptions at  $z$~=~3.538 along  two
sightlines  separated by only 13$h^{-1}$~pc.  However, as the velocity
difference  between the quasar and  the system is  only
6000~km~s$^{-1}$,  it cannot  be excluded  that the latter  system is
somehow associated  with the  quasar. 
Variations  of the strength  of
metal  line  systems  have also  been  reported along  adjacent lines
of sight with  larger separations (5--10~kpc) by Monier  et al. (1998)
and Lopez  et al. (1999). Each time however a damped Lyman-$\alpha$ 
system is seen along one of the sightlines. Contrary to  these previous 
studies, the Mg~{\sc ii} systems we examine here  are most likely to be 
associated with halos of intervening galaxies.

From  the  detection  of associated galaxies, radii of the order of 
35$h^{-1}$~kpc have been derived
for Mg~{\sc  ii} halos producing absorptions with equivalent widths 
$W_{\rm r}$~$>$~0.3~\AA~ at $z$~$<$~1
(Bergeron \& Boiss\'e 1991, Steidel 1993). 
Dimensions of the same order have been derived from
the study of Mg~{\sc  ii} systems seen along two  lines of sight
separated by  3~arcsec (Smette et  al. 1995).  The latter  authors find a
lower limit of 22~$h^{-1}_{50}$~kpc for  the radius of Mg~{\sc ii} absorbers
with $W_{\rm r}$~$>$~0.3~\AA~ at 0.5~$<$~$z$~$<$~1.3. 

If we assume that the lensing galaxy of APM~08279+5255 is at 
$z_{\rm lens}\sim 1$ (see next Section), the  separation between the  two 
lines  of  sight to A and B decreases from  1.7  
to  0.7~$h^{-1}_{75}$~kpc ($q_{\rm  o}$~=~0.5) between $z$~=~1.22  
and $z$~=~2.04 and is more than an order of magnitude smaller than the radius 
of Mg~{\sc ii} halos at intermediate redshift. 
Although evolution is probable,
it would be really surprizing that  the  Mg~{\sc   ii}  systems  studied  
here with  $W_{\rm r}$~$<$~0.5~\AA~  at $z_{\rm  abs}$~$>$~1.2  have 
characteristic  dimensions more than an order of magnitude smaller  
than what  is derived at lower redshift. If true, this would suggest  that the
structure of the Mg~{\sc ii}  halos at these redshifts differs substantially
from that at lower redshifts.  

A more likely explanation of these observations is that the halos are
composed of a collection of clouds (see Petitjean \& Bergeron 1990;
Srianand \& Khare 1994) and that individual clouds 
cover only one sight line. 
The number density of clouds is large enough so
that the total covering factor of the halo is close to one, consistent
with observations of associated galaxies.  However, individual clouds,
regularly spread over the velocity profile by kinematics,
cover only one image of the lens. The number density of clouds is not large 
enough
for the absorption material to cover the two lines of sight at all
velocities.  The distance over which the optical
depth, and hence the column density of \mgii, changes by at least 
one order of magnitude at \zabs$<$1.7 is smaller than $\sim1h^{-1}_{75}$ kpc.
In  contrast,  the two  strong  Mg~{\sc ii}  systems at  $z_{\rm
abs}$~=~1.06 and  1.18 (see  below) have covering  factor equal to  one (the
lines   are  saturated  and   go  to   the  zero   level)  over   more  than
200~km~s$^{-1}$.  These latter systems are likely to arise due to absorption
through central regions of  galaxies where  the  number of  clouds is  
so large  that saturated absorption occurs along both lines of sight 
whatever the radius of the individual clouds might be. 

\section {The two Mg~{\sc ii} systems at $z_{\rm abs}$~=1.062 and 1.181}
\begin{figure}
\centerline{\vbox{
\psfig{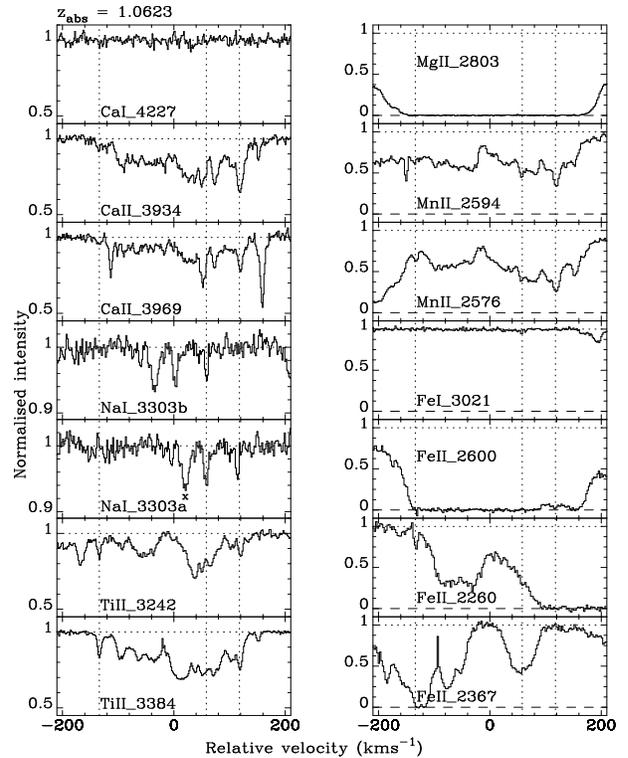}
}}
\caption[]{Absorption in a few transitions on a velocity 
scale with origin at $z_{\rm abs}$~=~1.06230. Vertical dashed lines 
mark velocity components discussed in the text.}
\label{106}
\end{figure}

\begin{figure}
\centerline{\vbox{
\psfig{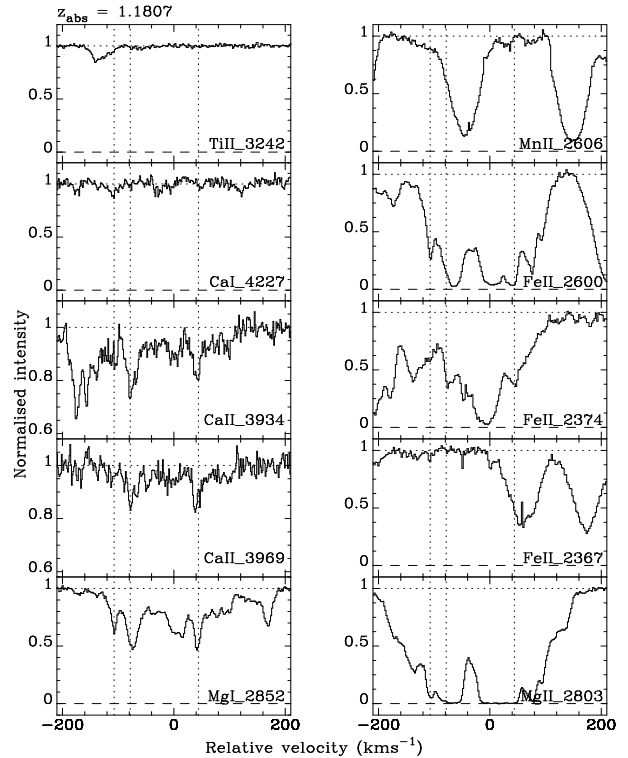}
}}
\caption[]{Absorptions in a few transitions on a velocity 
scale with origin at $z_{\rm abs}$~=~1.18070. Vertical dashed lines 
mark velocity components discussed in the text.}
\label{118}
\end{figure}

The    presence    of    a    strong    Mg~{\sc    ii}    system    ($W_{\rm
r}\lambda$2803~$\sim$~2.4~\AA)  at  $z_{\rm  abs}$~$\sim$~1.18  was  already
mentioned  by Irwin et  al. (1998).   There is  an additional  even stronger
Mg~{\sc    ii}    system    at    $z_{\rm    abs}    =    1.062$    ($W_{\rm
r}\lambda$2803~$\sim$~3.3~\AA).    Although  the   Mg~{\sc  ii}   lines  are
redshifted  in the  Lyman-$\alpha$  forest  and may  be  blended with  
Lyman-$\alpha$  intervening  absorptions, the  existence  of  the system  is
confirmed by numerous lines  redshifted redward of the quasar Lyman-$\alpha$
emission. As the two brightest images of the lensed quasar have similar 
magnitudes, it
is expected that the lines of sight to  both images pass through the core of
the  lensing  object  where  strong  Mg~{\sc ii}  absorption  is  likely  to
occur. The gravitational lensing may  thus result from the cumulative effect
of the two galaxies associated with these two absorbing systems
together with the objects responsible for the possible system at 
$z_{\rm abs}$~=~1.1727 (see Section~3.1) and the other damped 
Ly$\alpha$ system at $z_{\rm abs}$~=~2.974.
Absorptions from Mg~{\sc ii}, Fe~{\sc ii}, Ca~{\sc ii}, Mn~{\sc ii},
Ti~{\sc ii} and Na~{\sc i} are seen in both systems (see 
Figs.~\ref{106},\ref{118}). 
%
%
%
%
%
%
%
%
\subsection{Na~{\sc i} absorptions}
\begin{figure}
\vbox{
\centerline{
\psfig{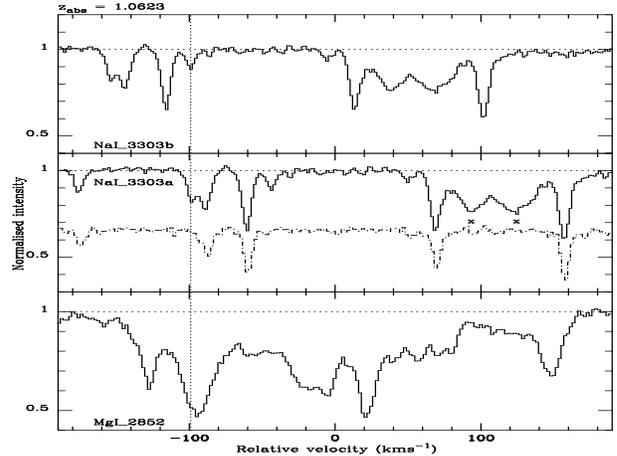}}
}
\caption[]{Portions of the APM~08279+5255 spectrum centered on the expected
positions of Na~{\sc i}$\lambda$3303.3
(top panel), Na~{\sc i}$\lambda$3303.9 (middle panel) and Mg~{\sc i}$\lambda$2852
(bottom panel) at $z_{\rm abs}$~=~1.181. The Na~{\sc i} component at 
$z_{\rm abs}$~= 1.1801, detected by the two lines of the doublet is indicated by a 
vertical dashed line. The spectrum of the standard star is overplotted in the
middle panel. Metal lines from other systems are indicated by crosses.
%
}
\label{na118}
\end{figure}

In both  Mg~{\sc ii}  systems, the weak  Na~{\sc i}$\lambda\lambda$3303,3303
doublet   is  detected  (see  Figs.~7   and \ref{na118}). 
The fact  that the two lines of  the doublet are
seen with  consistent strengths gives confidence that  the identification is
correct.  We have identified lines from other  metal line systems
in  the vicinity  of the  doublet.  The  spectra of  APM~08279+5255  and the
standard star are  compared in Fig.~\ref{na118} to rule  out the possibility
that the absorption features are of atmospheric origin.

The  column   densities  obtained  by  Voigt  profile   fitting  are  large,
log~$N$(Na~{\sc i})~=~12.9 and 13.5 at, respectively, $z_{\rm abs}$~=~1.0626
and  1.1801.   The  separation  of  the  two principal  lines  of  sight  is
$\sim$1.9$h^{-1}_{75}$~kpc at  $z$~$\sim$~1.  It  is therefore possible 
that the
clouds seen by  their Na~{\sc i} absorptions do not  cover the two brightest
lines of  sight.  In that  case, however, the  column density could  be even
larger by a factor of two (see Section~4.4).

Using the measurements by Sembach et al. (1993) and Diplas \& Savage (1994),
Bowen  et al.  (1995)  find that , in our Galaxy, 
log~$N$(H~{\sc i})~=~0.688~log~$N$(Na~{\sc i})~+~12.16.  Note that  this 
correlation holds  up to column
densities log~$N$(H~{\sc  i})~$>$~21 (see e.g.  Ferlet et  al.  1985).  
Applying this correlation for the $z_{\rm abs}$~=~1.0626 and  1.1801 absorbers
gives neutral hydrogen column densities of the order of log~$N$(H~{\sc
i})~$\sim$~21.0 and 21.4 respectively, 
for  gas-phase metallicities  comparable to  what is
seen in the  interstellar medium in our Galaxy  (the neutral hydrogen column
densities   could   be   even   larger   if   the   metallicity   in   these
intermediate-redshift  systems  is  smaller   than  in  our  Galaxy).   
Such large values for $N$(H~{\sc i}) are  supported by the column  densities 
of other  species that are
found to be  surprisingly close to what is  observed in typical interstellar
clouds (see below  and Table~2).  This leaves little  doubt that the systems
are indeed damped Lyman-$\alpha$ systems.

In our Galaxy, such high Na~{\sc  i} column densities are seen only in dense
and cool gas (see below).  The  typical $b$ values of the Na~{\sc i} diffuse
components  in both  the local  and  low-halo gas  is about  0.7~km~s$^{-1}$
corresponding to $T$~$<$~500~K (Welty et al. 1994).  Although the conclusion
is    very    uncertain   given    the    resolution    of   the    spectrum
($R$~$\sim$~37500) and the double  nature of the background source, the
lines we  observe are consistent with  $b$ values as  small as 1~km~s$^{-1}$
(see below).   From this,  we derive  an upper limit  on the  temperature of
$T$~$<$~2000~K.

%
%
%
\subsection{Comments on each system}
%
\subsubsection{$z_{\rm abs}$~=~1.062}
A  subset  of   the  absorptions  detected  in  this   system  is
shown  in Fig.~\ref{106}. It  can be  seen that
strong  (but unsaturated) absorptions from Ti~{\sc  ii} and Ca~{\sc ii}
are  detected.  The profile of the Ti~{\sc ii} absorption is spread
over about 250~km~s$^{-1}$ but does not  show any  edge  leading
pattern  (Prochaska  \& Wolfe  1998).  We  have
selected  to  examine two  components  at  $z_{\rm abs}$~=~1.0613  and
1.0631 because they  show well detached absorptions in  Ti~{\sc ii},
Mn~{\sc ii},  Ca~{\sc  ii}  and/or  Fe~{\sc   ii}  plus  the
component  at  $z_{\rm abs}$~=~1.0626 in  which we see Na~{\sc  i}.
Column densities  are listed in Table~2.
We have adjusted the best values for  $b$ from the fit to the lines that are
free of any  blending, considering for simplicity that  all species have the
same Doppler parameter and assuming complete coverage.  
We find $b$~=~1.5, 1.5 and  1.9 for, respectively,
$z_{\rm abs}$~=~1.0613, 1.0626 and 1.0631.

Ca~{\sc  i} is  not detected  and the  3$\sigma$ upper  limit on  the column
density in the  three components we have selected  is $<$~10.43, 10.43
and    10.20.     The     Mg~{\sc    ii}$\lambda$2796,2803    and    Fe~{\sc
ii}$\lambda\lambda\lambda$2382,2600,2586        lines        are       badly
saturated. Moreover, they are  redshifted in the Lyman$\alpha$ forest, which
prevents  any   fit  of  the   lines.   However,  the   unsaturated  Fe~{\sc
ii}$\lambda$2260 features detected at $z_{\rm abs}$~=~1.0613 and 1.0626 (see
Fig.~\ref{106}) give a reliable estimate of the Fe~{\sc ii} column density,
log~$N$(Fe~{\sc ii})~=~13.90$\pm$0.80 and 14.10$\pm$0.80 for 
$b$~=~1.5~km~s$^{-1}$, the continuum being adjusted locally. 
The non-detection of
Fe~{\sc  ii}$\lambda$2367 at  $z_{\rm  abs}$~=~1.0631 gives  an upper  limit
log~$N$(Fe~{\sc ii})~$<$~14.60.
\subsubsection{$z_{\rm abs}$~=~1.181}
A  subset  of   the  absorptions  detected  in  this   system  is  shown  in
Fig.~\ref{118}.  The profile  of the  Mg~{\sc i}  
absorption is
spread over  more than 200~km~s$^{-1}$ but,  as for the  previous system, it
does not  show any edge leading  pattern.  A number of  absorption lines are
optically thin or moderately saturated  and reliable column densities can be
derived even  though difficulties arise  from most of the  components being
badly  defined (see Fig.~\ref{118}).  We have selected for
study two subcomponents  which are clearly seen in  all absorption profiles
at $z_{\rm abs}$~=~1.1799 and 1.1801.
They are indicated on  Fig.~\ref{118} by vertical dashed 
lines, and the  column densities obtained  from Voigt-profile fitting are  
given in Table~2.  Doppler parameters  have been
considered to be identical for all species.
For $z_{\rm abs}$~=~1.1799 we find $b$~=~2.5~km~s$^{-1}$.

For the component  at $z_{\rm abs}$~=~1.1801 in which  Na~{\sc i} absorption
is detected, the Doppler parameter  is estimated by fitting the well-defined
lines of  the sodium doublet after  having taken  into account the
effect  of  Na~{\sc  i}$\lambda$3303.3  being  partially  blended  with  an
atmospheric  feature   (see  Fig.~\ref{na118}). We  obtain   a  best  value
$b$~=~1.1$^{+1.0}_{-0.5}$~km~s$^{-1}$. 
The column densities derived using the two values $b\pm$~1$\sigma$ differ by
large factors. We  have therefore refined the determination of $b$ and $N$ 
using the following indirect argument.

The ratio of  the Mg~{\sc i}  to the Na~{\sc i}  column densities in neutral
gas can be written
\begin{equation}
{N({\rm Mg~{\sc i}})\over N({\rm Na~{\sc i}})} =
{{\rm Mg~{\sc i}}\over {\rm Mg}} \times 
{{\rm Na}\over {\rm Na~{\sc i}}} \times {\delta_{\rm Mg}\over 
\delta_{\rm Na}}  \times {Z_{\rm Mg}\over Z_{\rm Na}},
\end{equation}
where $\delta$ is the  depletion of the element  due to the presence of dust
and $Z$ the abundance. Assuming that (i)  the relative abundance of Na to
Mg   is solar, $Z_{\rm Mg}$/$Z_{\rm  Na}$~=~19,  (ii) the relative depletion
into dust-grains is  $\delta_{\rm Mg}$/$\delta_{\rm  Na}$~$>$~0.3 (Savage \&
Sembach 1996) and (iii) (Mg~{\sc i}/Mg~{\sc ii})$\times$(Na~{\sc ii}/Na~{\sc
i})~$>$~0.15 in  cold and neutral gas (P\'equignot \& Aldrovandi 1986),  we
derive $N$(Mg~{\sc  i})/$N$(Na~{\sc  i})~$\geq$~0.8.   This   estimation  is
certainly  very  approximate. However, this simple   argument shows that the
latter   ratio cannot    be  much  smaller    than  0.5.    If we   assume
$b$~=~1.5~km~s$^{-1}$, then  we find $N$(Mg~{\sc  i})/$N$(Na~{\sc i})~=~0.045
which  is definitively too small. We  therefore have  fitted the absorption
lines decreasing $b$ from 1.5 to 0.5~km~s$^{-1}$ to find the largest
$N$(Mg~{\sc i})/$N$(Na~{\sc i}) ratio. We find a maximum
$N$(Mg~{\sc i})/$N$(Na~{\sc i})~=~0.3 for $b$~=~0.8~km~s$^{-1}$,
log~$N$(Mg~{\sc i})~=~13.0 and log~$N$(Na~{\sc i})~=~13.5.

\subsection{Physical state of the gas}
\begin{table*}
\begin{tabular}{llllllll}
\multicolumn{8}{l}{{\bf Table 2.} Column densities$^a$}\\ 
\hline
\multicolumn{1}{c}{Redshift}&
\multicolumn{1}{c}{1.0613}&\multicolumn{1}{c}{1.0626}&
\multicolumn{1}{c}{1.0631}&
\multicolumn{1}{c}{1.1799}
&\multicolumn{1}{c}{1.1801}&\multicolumn{1}{c}{23 Ori$^b$}
&\multicolumn{1}{c}{$\mu$Col$^d$}\\
\hline\\
H~I & & & & & & 20.74 & 19.86 \\
H$_2$ & & & & & & 18.30 & 15.50 \\
Na~I   & $<$11.90 & 12.91$\pm$0.04 & $<$12.80 & $<$12.40 & 13.50$\pm$0.11 & 
13.36 & 11.6$^e$ \\
Mg~I   & 11.6:$^c$ & bl & bl & 11.61$\pm$0.06 & 13.0$\pm$0.74 & 
13.81 & 12.55 \\
Mg~II  & & & & & & 15.68 & 15.08 \\
Ca~I   & $<$10.43 & $<$10.43 & $<$10.20 & $<$10.80 & $<$10.33 & 
10.20 & \\
Ca~II  & 11.31$\pm$0.11 & bl & 11.71$\pm$0.03 & 11.33$\pm$0.39 &
11.79$\pm$0.35  & 12.10 & 12.19 \\
Ti~II  & 11.80$\pm$0.05 & $<$12.00 & 11.90$\pm$0.02 & $<$11.66 & $<$11.20 
& 11.23 & 11.78 \\
Mn~II  & bl & bl & 12.70$\pm$0.40$^c$ & bl & bl & 13.15 & 12.48 \\
Fe~I   & $<$11.50 & $<$11.63 & $<$11.40 & $<$11.40 & $<$11.50 & 
11.34 & \\
Fe~II  & 13.90$\pm$0.80 & $<$14.10 & $<$14.60 & 13.36$\pm$0.07 
& $<$15.00 & 14.38 & 14.13 \\
CH+    & $<$12.80 & $<$12.70 & $<$13.00 & $<$13.20 & $<$13.30 &
13.06 & \\
CH     & $<$13.40 & $<$13.60 & $<$13.50 & $<$13.40 & $<$13.60 &
12.69 & \\ 
\hline 
\multicolumn{8}{l}{$^a$ logarithm of, in cm$^{-2}$; Doppler parameters
are taken to be $b$~=~1.5, 1.5, 1.9, 2.5, 0.8~km~s$^{-1}$ for 
the five }\\
\multicolumn{8}{l}{components respectively; the components are assumed 
to cover all images;}\\  
\multicolumn{8}{l}{$^b$Welty et al. (1999); $^c$The continuum is fitted 
locally; $^d$ Howk et al. (1999); $^e$ Hobbs (1978).}\\    
\multicolumn{8}{l}{A sign "bl" means that no measurement is possible due to
blending effects} \\
\label{coldens}
\end{tabular}
\end{table*}
Table~2 contains  the column densities   measured in the five  subcomponents
defined  above.   The  penultimate  column gives for   comparison the column
densities  measured   by Welty et  al.   (1999)  in the   neutral gas toward
23~Ori. This gas is   found to have temperature   $T$~$\sim$~100~K, hydrogen
density $n_{\rm H}$~$\sim$~10--15~cm$^{-3}$ (and therefore total thickness
of        12--16~pc)        and         electronic      density      $n_{\rm
e}$~$\sim$~0.15$\pm$0.05~cm$^{-3}$.   The  last column of  Table~2 gives the
measurements  for the strongest  component of the warm neutral gas
toward $\mu$Col  (Howk \&  Savage 1999). This    gas is found   to have
$T$~$\sim$~6000--7000~K and $n_{\rm  e}$~$\sim$~0.3~cm$^{-3}$.  The two sets
of column  densities  are similar except  for $N$(Na~{\sc  i}) which is much
smaller toward $\mu$Col. This illustrates directly that at least in the
component at $z$~$\sim$~1.1802 toward APM~08579+5255 where
Na~{\sc i} is detected, the gas is most likely to be 
neutral    and cold.  Moreover,   note also  that   in  our  Galaxy, a ratio
$N$(Na~{\sc  i})/$N$(Ca~{\sc     ii})~$>$~1,    as   observed   at   $z_{\rm
abs}$~=~1.1802  toward APM~08579+5255, is characteristic of  cold gas in the
disk. Indeed,  along the line of   sight to the LMC,   such large ratios are
observed only at the systemic  velocities of the LMC and  the Galaxy; gas in
between   has  $N$(Na~{\sc  i})/$N$(Ca~{\sc   ii})~$<$~1  (Vidal-Madjar   et
al. 1987, Vladilo et al. 1993).


From the  upper limit on  Ca~{\sc i} and  Fe~{\sc i} we  can  derive, in the
components where $N$(Ca~{\sc ii}) and $N$(Fe~{\sc ii}) are measured, a lower
limit for the electronic density for a given ionizing field. Indeed, $n_{\rm
e}$~=~($X^{\rm o}$/$X^{\rm  +}$)$\times$($\Gamma$/$\alpha$), 
where  $\Gamma$ is  the
photoionization  rate  and  $\alpha$  the   recombination coefficient.   The
corresponding  electronic   densities  for    a  Galactic    ionizing  field
(P\'equignot \&  Aldrovandi 1986), are  $n_{\rm e}$~$<$~0.13 and 3~cm$^{-3}$
for Fe and  Ca at $z_{\rm abs}$~=~1.0626  and 1.1801 respectively.  
Note that the  determination  of $n_{\rm e}$  in the
interstellar medium from  the ratio of singly ionized  to neutral species is
highly uncertain probably because of contamination of  the singly ionized
column density determination by adjacent components (Welty et al. 1999).
Writing the same relation for sodium and equating the 
expression of $n_{\rm e}$ obtained for sodium and iron or calcium
leads to upper limits on the Na~{\sc i}/Na~{\sc ii} ratio which depends only
on the shape of the  ionizing spectrum and not on  its absolute value.  With
the only assumption that the ionizing spectrum has the  same shape as in our
Galaxy  and  using the coefficients derived   from P\'equignot \& Aldrovandi
(1986; see Welty et al. 1999), we find log~Na~{\sc i}/Na~{\sc ii}~$<$~$-$1.3
and 0.1 from the constraints obtained on Fe and  Ca in the 1.0626 and 1.1801
systems respectively.   

From $N$(H~{\sc i})~=~$N$(Na~{\sc i})$\times$(Na/Na~{\sc i})/$Z$(Na)
and using the two upper limits on the Na~{\sc i}/Na~{\sc ii} ratios 
derived above, we can write
$N$(H~{\sc i})~$>$~19.9 and 19.5/($Z$(Na)/$Z_{\odot}$(Na))/$\delta$(Na) 
at $z_{\rm abs}$~=~1.0626  and 1.1801 respectively,
where  $\delta$(Na) is  the fraction of   sodium remaining in the gas  phase
after depletion into  dust-grains. This factor is equal  to about 0.1 in the
ISM (Savage \& Sembach 1996).
This adds  support to arguments presented  previously that these systems are
damped.
To illustrate the discussion, simple photo-ionization  models using the code
Cloudy (Ferland 1996)   have been  constructed.    The absorbing  cloud   is
modelled  as a  plane parallel  slab  with uniform  density, solar  chemical
composition and neutral  hydrogen column density $5\times10^{20}$~cm$^{-2}$.
The elements are considered to  be depleted into dust-grains  as in the cool
cloud  observed toward $\zeta$Oph  (Savage \& Sembach 1996).  The shape
of   the  UV       flux    is       taken   to    be       a       power-law
$F_{\nu}$~$\propto$~$\nu^{-1.0}$. The  resulting column densities of various
species along a line-of-sight perpendicular to the slab are given versus the
ionizing parameter in Fig.~\ref{model}.  As discussed above, the $N$(Mg~{\sc
i})/$N${(Na~{\sc i}) ratio can be smaller than one only  if magnesium is more
depleted into dust grains than sodium. Note that every model that produces
enough Na~{\sc i} has temperature less than 100~K.
\begin{figure}
\centerline{\vbox{
\psfig{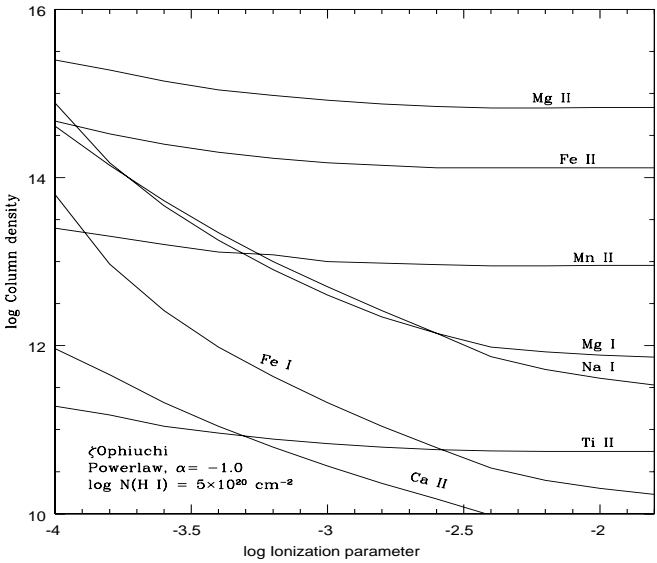}
}}
\caption[]{Results of photoionization models with $Z=Z_\odot$,
log~$N$(H~{\sc i}) = 20.7, constant density and plane parallel
geometry. Depletion into dust-grains is assumed to be the same as
in the cool cloud toward $\zeta$Oph (Savage \& Sembach 1996).
Column densities are given versus ionization parameter; the shape
of the UV flux is a power-law $F_{\nu}$~$\propto$~$\nu^{-1.0}$.
}
\label{model}
\end{figure}
\par\noindent
Finally, we do not detect any CH$\lambda$4300 and  CH$^+\lambda$4232
absorption. The limits on the column densities are log~$N$(CH) $<$ 13.5 and
log~$N$(CH$^+$) $<$ 13.0  at both   $z_{\rm  abs}$~=~1.06  and    1.18  (see
Table~2).  This  is just  what  would be  expected in  our  Galaxy  along an
otherwise similar line of sight. Indeed, along the  line of sight to 23~Ori,
log~$N$(CH$^+$)~=~13.06 and  log~$N$(CH)~=~12.69 (Welty   et al.  1999;  see
Table~2). More  generally, the column density  of CH is observed to increase
from 1.5  to 7.5$\times$10$^{13}$~cm$^{-2}$ for  lines of sight with $E_{\rm
B-V}$ increasing from 0.5 to 1.5 (Gredel et al. 1993).  It would be of prime
interest to obtain better data in  this wavelength range to better constrain
the molecule column densities.

\subsection{Consequence of partial covering factor}
From the  detection of the  Na~{\sc i}$\lambda\lambda$3303.3,3303.9 doublet,
we can derive  that the two components at  $z_{\rm abs}$~=~1.0626 and 1.1801
arise in  cold, dense  and neutral gas  (see Section~4.3).  It  is therefore
possible that the dimension of the cloud is less than $\sim$1.9~kpc which is
the separation of the lines of sight to the two brightest images at the 
redshift
of the  absorber assuming that the  lensing object is at  the same redshift.
Indeed,  large variations of  Na~{\sc i}  column density  have been
reported in  the nearby interstellar medium  on very small  scales (Meyer \&
Lauroesch 1999).  We have  therefore investigated the  impact on  the column
density  measurements of  the assumption  that  the cloud  covers the  three
images. If it is  the case that the cloud covers only  one of the images the
column density is larger. The  discrepancy cannot be very large, however, as
most  of the  lines used  for column  density determination  are  weak.  The
strong lines  are completely  saturated and blended,  which prevents  in any
case any determination of the line parameters.
\par\noindent
We  have considered  the  Mg~{\sc i} and  Na~{\sc  i} lines   in the $z_{\rm
abs}$~=~1.1801 component. It can be seen on Fig.~\ref{118} that if only one
image is covered, it cannot be the brightest as the residual normalized flux
in the Mg~{\sc i} absorption is smaller than 0.5.  As the  flux ratio of the
two brightest components is 1.2, we artificially placed the zero level
at 0.4 on the scale of Fig.~\ref{118}. Voigt profile fitting of the Na~{\sc
i} doublet gives log~$N$(Na~{\sc i})~=~13.7 and $b$~=~1~km~s$^{-1}$ which is
within a factor  of two of  what has been  derived previously (see Table~2).
Four components have  been used  to  fit the Mg~{\sc   i} blend at   $\Delta
v$~=~$-$60~km~s$^{-1}$ (see Fig.~\ref{118}).  For  the component at $z_{\rm
abs}$~=~1.1801, we obtain log~$N$(Mg~{\sc i})~=~13.8. \par\noindent
Note that in this case, the  Na~{\sc i} and  Mg~{\sc i} column densities are
nearly  identical to what is  observed  toward 23~Ori  (see Table~2). We
therefore conclude that $N$(Mg~{\sc i})/$N$(Na~{\sc i})~$\sim$~1 is a robust
measurement in  this system. 
\par\noindent 
Mg~{\sc i}  is  the line  with  the largest saturation  among  those used to
derive column densities quoted in Table~2.
Therefore, the column densities derived from weaker  lines should not differ
from what is quoted in Table~2 by more than a factor of two.
\subsection{Metallicity and dust content}
%
%
In the following we use the conventional definition  
[X/H] = log($Z/Z_{\odot}$), with $Z$(X) the
metallicity of species X relative to hydrogen.
Ca~{\sc  ii} and Ti~{\sc ii}  are both detected at $z_{\rm abs}$~=~1.0613 
and 1.0631. The   column densities are  consistent
with   what  is seen  in   the interstellar  medium   of  our Galaxy (Stokes
1978; see  Table~2). However, 
log~$N$(Ca~{\sc ii})/$N$(Ti~{\sc  ii})~$\sim$~$-$0.5  and  $-$0.2 at, 
respectively, $z_{\rm abs}$~=~1.0613 and 1.0631 when the relative solar
metallicity is log~$Z_{\odot}$(Ca)~$-$~log~$Z_{\odot}$(Ti)~=~1.38.
Various explanations for this discrepancy can be invoked, amongst
them the most likely are: (i) Calcium is mostly in the form of Ca~{\sc
iii}, (ii) Calcium is more  depleted into dust-grains than   Titanium. 
Note that the relative metallicities [Ca/Fe] and [Ti/Fe] are
both observed to be $\sim$+0.3  for [Fe/H]~$<$~$-$1
in late-type stars (Th\'evenin 1998). 
Note also that
the Ca~{\sc iii}/Ca~{\sc  ii} ratio  derived in  our Galaxy is  in the range
5--10 which is much smaller than the discrepancy mentioned above.
This favors the explanation that Calcium is heavily depleted into
dust-grains (see below). 
\par\noindent 
In  the $z_{\rm  abs}$~=~1.0631 component,  Mn~{\sc ii}  is also  seen
and  log~$N$(Mn~{\sc ii})/$N$(Ti~{\sc ii})~$\sim$~$+$0.8. As
the solar metallicity of Mn is  $-$6.47, the relative solar metallicity 
is  log~$Z_{\odot}$(Mn)~$-$~log~$Z_{\odot}$(Ti)~=~~=~$+$0.6.   
This  is  consistent  with  
similar  depletion of  Mn  and Ti  as observed  in warm  gas
(Savage \&  Sembach 1996).  
Therefore, in this component,  we cannot  rule out that  the low
Ca~{\sc ii} column density is due to ionization.
\par\noindent 
In  the $z_{\rm  abs}$~=~1.0613 component,  Fe~{\sc ii}  is also  seen
with log~$N$(Fe~{\sc  ii})/$N$(Ti~{\sc ii})~$\sim$~$+$2.1.
The solar  metallicity of iron  and titanium are, respectively,  $-$4.49 and
$-$7.07 and the relative solar metallicity is 
log~$Z_{\odot}$(Fe)~$-$~log~$Z_{\odot}$(Ti)~=~$+$2.58. There is no
differential ionization correction for these two elements.
The    discrepancy, $\sim$+0.5~dex, between  the two    ratios  can only  be
explained   by   a larger  depletion   of   titanium  compared  to iron into
dust-grains by $\sim$0.5~dex   as in the   cool  gas of the   ISM (Savage \&
Sembach 1996).  Indeed, from nucleosynthesis alone, we would expect titanium
to be enhanced  compared  to  iron (Th\'evenin  1998)  contrary to   what is
observed.  We therefore conclude that depletion into dust-grains is
present in this system.
The low Ca~{\sc ii}  column density can indeed be  explained by a  large
depletion of calcium into dust-grains as is observed in the ISM. 
\par\noindent
The fact that Ti~{\sc ii} is not detected in the $z_{\rm abs}$~=~1.18 system
is surprising, although the limit on the  column density is not stringent and
only a factor of four smaller than what is  seen in the $z_{\rm abs}$~=~1.06
components.
\par\noindent
The presence of dust is supported by the analysis
of the column density ratios in
the $z_{\rm abs}$~=~1.0626 and 1.1801 components where Na~{\sc i} absorption
is detected.    Indeed, we can compute  for  different elements the quantity
$N_{\odot}$~=~10$^{[X/H]}$$\times$$N$(H~{\sc i}) which is the H~{\sc i} 
column density of a cloud with solar metallicity that would have the
same column density of element $X$ as the observed one. 
We compute,
\begin{equation}
{\rm log} N_{\odot} = [{\rm X/H}] + {\rm log} N({\rm HI}) = 
{\rm log} \biggl( N({\rm X}^i) {{\rm X}\over {\rm X}^i} 
{1\over\delta} {1\over Z_{\odot}({\rm X})}\biggr) 
\end{equation}
where $Z_{\odot}$  is the solar  metallicity and 1-$\delta$ the  fraction of
the element tied  up into dust-grains.  We assume  solar metallicity and 
depletion  pattern given  in Table~5  of  Savage \&  Sembach (1996).   Solar
metallicities relative to hydrogen for  Na, Fe, Ca and Ti are, respectively,
$-$5.69,  $-$4.49, $-$5.66  and $-$7.07.  If we  assume that  the absorption
arises in cool  gas similar to the gas seen in  front of $\zeta$Oph, we
find  that at $z_{\rm  abs}$~=~1.1801, log~$N_{\odot}$~$>$20.44,  
$<$21.80, $<$21.18, $<$21.30  from   Na,  Fe,   Ca  and  Ti   and  at   
$z_{\rm  abs}$~=~1.0626,
log~$N_{\odot}$~$>$20.85, =~20.86, $<$22.10 from Na, Fe and Ti respectively.

These consistent results suggest that,  in  the  $z_{\rm  abs}$~=~1.0626 and  1.1801
components,  log~$N$(H~{\sc   i})~+~[X/H]~$\sim$~21 and the  depletion into
dust-grains is similar to what is seen in cool Galactic interstellar clouds.
If the relation   of Bohlin et al. (1978)   holds, this implies a  color
excess $E_{\rm    B-V}$~$\sim$~0.2  with  the   only  assumption   that  the
overall dust-to-metal ratio does not depend on metallicity. At the wavelength 
of the  Mg~{\sc   ii}  absorption,   ($\lambda_{\rm  z=1}$~$\sim$~3000~\AA;
$\lambda^{\rm QSO}_{\rm rest}$~$\sim$~1200~\AA;
$\nu^{\rm QSO}_{\rm rest}$~$\sim$~2$\times$10$^{15}$~Hz)
this would induce an extinction of about 1~mag (Seaton 1979). At 
$\nu^{\rm QSO}_{\rm rest}$~$\sim$~10$^{14}$~Hz, 
the extinction would be negligible.  
Note that there is  some  evidence that $\nu F_{\nu}$
decreases  by   a factor   of two  from  10$^{15}$ to   10$^{14}$~Hz in  the
APM~08279+5255 SED (Lewis et al. 1998).

The amount of dust suggested by the  previous discussion is significant.  We
have therefore searched  the spectrum of  the  quasar for some signature  of
this  amount  of  dust.   For  this,  we   have  compared  the  spectrum  of
APM~08279+5255   with the composite QSO  spectrum  obtained with the FOS-HST
(Zheng et al. 1997) attenuated by dust with optical depth 
$\tau_{\rm dust}$($\lambda$) at the observed wavelength $\lambda$,
\begin{equation}
\tau_{\rm dust}(\lambda)~=~k \bigg[ {N \over10^{21}{\rm cm^{-2}}}\bigg]\xi\bigg(
{\lambda\over1+z} \bigg)
\end{equation}
where $\xi(\lambda)$  is  the ratio  of the   extinction at the   wavelength
$\lambda$  to that in the  B-band as observed in our Galaxy, 
$k$~=~10$^{21}$~cm$^{-2}$~$\tau_{\rm B}$/$N$(H~{\sc i})
is  the dimensionless dust-to-gas
ratio and $N$ the H~{\sc i} column density (e.g. Pei et al. 1991,
Srianand \& Kembhavi 1997). We assume here log~$N$(H~{\sc i})~=~21.
The redshift of the QSO is taken to be $z_{\rm QSO}$~=~3.91 
and  that of  the absorber $z_{\rm   abs}$~=~1.1. The results  are shown  in
Fig.~\ref{dust}  where the spectrum  of APM~08279+5255 as observed over part
of the $R$-band (solid line) is plotted together  with the composite HST-FOS
QSO spectrum attenuated by dust with  optical depth in the B-band
$\tau_{\rm dust}$(B)~=~0.1,
0.2 and 0.3.   It is apparent that the  best fit is obtained with $\tau_{\rm
dust}$(B)~$\sim$~0.3. In our Galaxy, this would correspond to 
about $E_{\rm B-V}$~$\sim$~0.1. This is two times smaller than 
what has been derived above. This suggests that the dust to metal ratio
in the redshifted gas is about half that in our Galaxy.
 
Altogether we  find that the H~{\sc i}  column density at $z$~=~1  is of the
order  of $1\times  10^{21}$~cm$^{-2}$ to  $5\times  10^{21}$~cm$^{-2}$, the
corresponding  metallicity  is  in  the  range  1--0.3~$Z_{\odot}$, the
dust-to-metal ratio is about half that in our Galaxy and the relative
depletion of species into dust-grains is  similar to what is observed in 
cool  gas in the disk of our Galaxy.

The colors of the images derived from HST imaging are nearly identical 
(Ibata et al. 1999). The differential reddening over kpc scales is smaller 
than 10\%. This indicates that, although the medium is highly inhomogeneous,
the extinction is fairly uniform over 
distances of the order of the separation between the lines of sight.
It must be noted that the number of components in the two strong Mg~{\sc ii}
systems must be quite large. Indeed, the Mg~{\sc ii} absorptions 
reach the zero level over $\sim$200 and 350~km~s$^{-1}$ at, respectively,
$z_{\rm abs}$~=~1.181 and 1.062. One single component cannot be much larger 
than about $\sim$10~km~s$^{-1}$ and therefore the number of components is 
larger than 20 and 40 respectively along all three lines of sight.
This is probably larger than what is seen in the disk of our Galaxy 
(Sembach \& Danks 1994). Therefore, it may well be possible that the total 
reddening we see is not due to one strong component only but rather 
is due to the accumulated effect of a large number of diffuse clouds
with small and similar extinctions. As the number of clouds is large and 
similar along the different lines of sight, the differential extinction 
is small. To probe this, higher S/N ratio data should be obtained in the 
region of the Na~{\sc i} absorption to investigate what is the velocity 
spread of this absorption. 
\begin{figure}
\centerline{\vbox{
\psfig{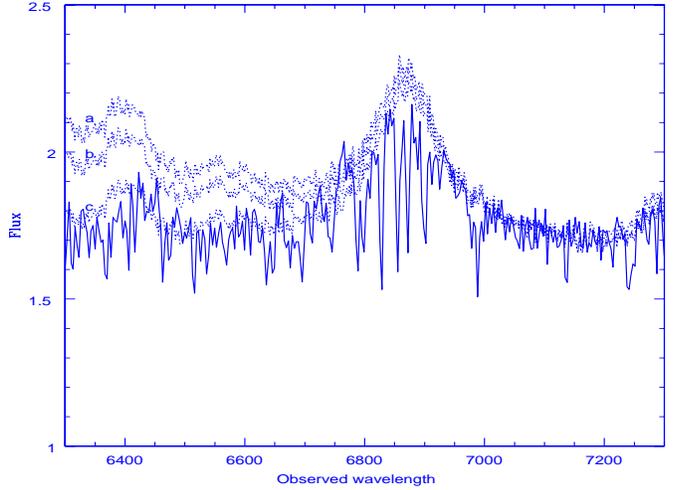}
}}
\caption[]{Spectrum of APM~08279+5255 as observed over part of the $R$-band 
(solid line) plotted together with the composite QSO spectrum obtained
from FOS-HST observations (Zheng et al. 1997) and attenuated by dust with 
optical depth in the B-band, $\tau_{\rm dust}$(B)~=~0.1 (a), 0.2 (b) and 0.3 
(c; see text). The Galactic extinction curve is assumed.}
\label{dust}
\end{figure}
\subsection{Nature of the systems}
Churchill (1999) has shown that Mg~{\sc ii} systems at intermediate redshift
can be   classified  in five categories:  DLA,   Double, Classic, C~{\sc iv}
deficient and  Weak. The first class is  characterized by strong Mg~{\sc ii}
absorption saturated  over   $\sim$150~km~s$^{-1}$ and, when  observed,  the
Lyman-$\alpha$  line is  damped.  The Double  systems  are characterized  by
kinematic velocity spreads up  to 400~km~s$^{-1}$. Other  classes correspond
to much weaker systems.

The  system  at $z_{\rm  abs}$~=~1.181 would be   classified  as Double (see
Fig.~\ref{118}): it  has Mg~{\sc ii}$\lambda$2796 absorption saturated over
more than  150~km~s$^{-1}$  in total. The  characteristic  double profile is
very similar   to that  of   the  system   at  $z_{\rm  abs}$~=~1.17  toward
Q~0450--132   (see  Petitjean  et  al.   1994).    The  system    at $z_{\rm
abs}$~=~1.062 would be classified  as DLA, as the  Mg~{\sc ii} absorption is
continuously     saturated    over    more   than     300~km~s$^{-1}$   (see
Fig.~\ref{106}). To our  knowledge, this  system has  one of the  strongest
Mg~{\sc ii} absorption features known ($W_{\rm r}$~$\sim$~3.3~\AA).
As seen from  Fig.~\ref{106},  the Ti~{\sc  ii} absorption
spans  $\sim$~300~km~s$^{-1}$ and coincides exactly  with the saturated part
of the Mg~{\sc ii} absorption.

At  high-z,  Prochaska  \&  Wolfe   (1998) have shown     that most  of  the
low-ionization absorptions associated   with   DLAs, have   an  edge-leading
profile, possibly revealing large-scale rotational motions. They conclude that
DLA systems arise  in large  rotating disks.   Haehnelt et al.  (1998)  have
claimed that the observed profiles can be reproduced as well  if the line of
sight passes through several interacting blobs. Indeed, Ledoux et al. (1998a)
have shown that the observed profiles are  consistent with rotation when 
they span less than $\Delta V$~$\sim$~150~km~s$^{-1}$. For larger velocity 
spreads, several sub-systems are usually seen.

The characteristic edge-leading profile is seen in the $z_{\rm abs}$~=~2.974
system  (see  below,  Fig.~\ref{s297})  but not  in  the  $z_{\rm
abs}$~=~1.062  and 1.181  Mg~{\sc  ii}  systems.  The  large  spread of  the
profiles ($>$~200  and 300~km~s$^{-1}$) is  more reminiscent of  the profile
seen toward the  supernova SN~1993J in the large  nearby spiral galaxy M~81.
The  latter is  part of a complex
interacting  group together with  M~82, NGC~3077  and NGC~2976  with tidally
stripped   H~{\sc  i}   linking  individual   galaxies  over   an   area  of
$\sim$50$\times$100~kpc$^2$ (Yun et al. 1994). The Mg~{\sc ii} absorption is
characteristic  of the  Double class  as defined  by Churchill  (1999).  
It is  spread over $\sim$400~km~s$^{-1}$ with two strong
saturated absorptions of width, respectively,  $\Delta V$~$\sim$~150 and
90~km~s$^{-1}$, and separated by  $\sim$180~km~s$^{-1}$ (Bowen et al. 1995).
Most of the absorption is due to tidal debris expelled outside
the disks of the interacting galaxies.

In the case of the $z_{\rm abs}$~=~1.062  system, the Mg~{\sc ii} absorption
does  not  show any sub-structure.    Such strong absorption  is expected to
occur in the central part of galaxies. Although the  statistics are very poor,
it  seems that when  strong absorption  occurs, the  equivalent width of the
absorption is anti-correlated with the impact  parameter between the line of
sight and the center of the galaxy (Bowen et al. 1995). 
The impact parameter could be as low as 1~kpc for a system with 
$W_{\rm r}$~$\sim$3~\AA.
The separation  between the    two    bright  images  of   APM~08279+5255 is
$\sim$1.9$h^{-1}_{75}$~kpc at $z$~$\sim$~1. This suggests that the object
giving rise to the $z_{\rm abs}$~=~1.06 Mg~{\sc ii} system could be 
nearly exactly aligned with the quasar.


%
\section{The system at $z_{\rm abs}$~=~2.974}
\begin{figure}
\centerline{\vbox{
\psfig{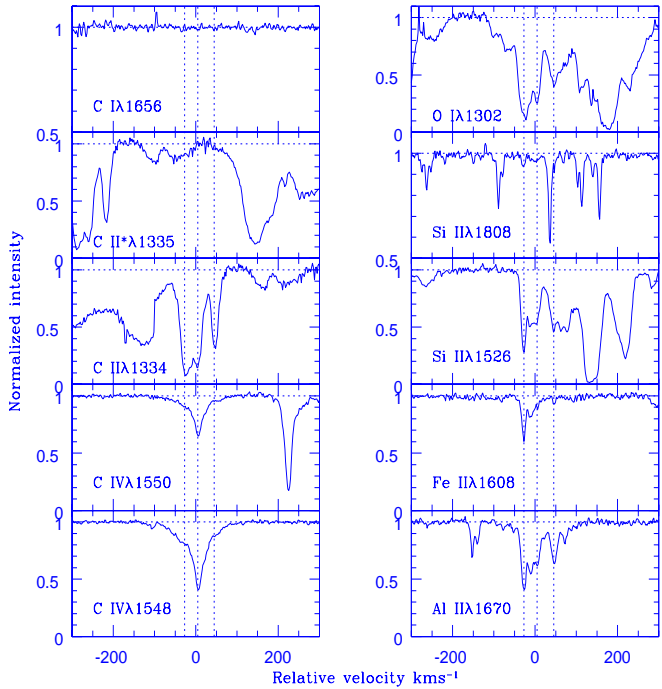}
}}
\caption[]{Absorptions in a few transitions on a velocity 
scale with origin at $z_{\rm abs}$~=~2.974. Vertical dashed lines 
mark the position of the strongest components.}
\label{s297}
\end{figure}
There  is   a strong absorption feature  with   $W_{\rm obs}$~$>$~19~\AA~ at
$\lambda$~=~4831~\AA. It corresponds  to H~{\sc i} Lyman-$\alpha$ at $z_{\rm
abs}$~=~2.974. The only possibility of coincidence with a BAL transition
could be Lyman$\beta$ at $z_{\rm abs}$~=~3.71 but there is no
corresponding Lyman$\alpha$ transition (see Srianand \& Petitjean 2000).
Although   the    red-wing  of  the  absorption     has  the
characteristic shape of a  damped transition, uncertainties in the continuum
determination   prevent   an  accurate    determination  of   the    column
density. Associated absorptions from Al~{\sc  ii}, Fe~{\sc ii}, Si~{\sc ii},
C~{\sc  ii} and   O~{\sc  i}   are  detected  in four    components spanning
$\sim$100~km~s$^{-1}$ (see Fig.~{\ref{s297}).  By fitting the Lyman-$\alpha$
line, we  estimate that  the total  H~{\sc i}  column   density in  the four
components is in the range 19.8~$<$~log~$N$(H~{\sc i})~$<$~20.3. 

It is important  to note that there  is no evidence that the  cloud does not
cover the three lines  of sight which are separated by
$\sim$~200$h^{-1}_{75}$~pc
at  the  redshift of  the  absorber.   Indeed, the  core  of  the H~{\sc  i}
absorption is  black over  $\sim$~15~\AA.  Column densities  integrated over
the  absorption profiles have been obtained for all species and summarized in
Table~3.   Column  \#3  of  Table~3  gives  the  abundance  of  the  element
assuming that log~$N$(H~{\sc i})~=~20.3 in  the cloud and that the 
observed ion is the dominant species.  Given the uncertainty in the
neutral hydrogen column density, the absolute values are 
unreliable and could be 0.5~dex higher.
Column \#4 and \#5
of Table~3 give the solar  metallicity and the metallicity relative to solar
respectively.  It  is remarkable how consistent the  measurements are, which
point  toward  metallicities   less  than  10$^{-1.5}$~$Z_{\odot}$.  The  low
metallicities  are  not due  to  depletion  into  dust-grains.  Indeed,  the
relatively  small  neutral  hydrogen  column  density  implies  that  column
densities of relatively  abundant elements can be measured.   It is apparent
that iron is  not depleted compared to carbon or oxygen  and that the amount
of dust in this cloud must be very small.

Metallicity in damped Lyman-$\alpha$ systems is usually measured using zinc,
an element  that is not  very much depleted  into dust-grains in  our Galaxy
and, because it  has relatively low metallicity compared  to other elements,
induces non-saturated  absorptions even for  clouds of high  hydrogen column
density.  There is barely no evolution in the measured Zinc metallicity from
$z$~$\sim$~1 to $z$~$\sim$~3 (Pettini et  al. 1997, 1999).  At $z$~$>$~3, in
most damped  Lyman-$\alpha$ studied  up to now,  zinc is not  detected.  
This is most probably a consequence of limited S/N ratio of the 
data however. Indeed, the detection limit of most of the spectra is
log~$N$(Zn~{\sc ii})~$\sim$~11.5 which means [Zn/H]~$<$~$-$1.4 (see e.g. 
Prochaska \&  Wolfe 1997). It  can be noted
that in the system at $z_{\rm abs}$~=~2.974 toward APM~08279+5255, the limit
on zinc is of this order.  However, as we can measure metallicities for more
abundant elements, we know that  metallicities are less than $-$1.5.  For such
abundances,   zinc  would   have  been   detectable  in   the   spectrum  of
APM~08279+5255   only   for  H~{\sc   i}   column   densities  larger   than
10$^{21}$~cm$^{-2}$.  It is therefore possible that the  upper limit found
for the zinc metallicity at $z$~$>$~3 in previous surveys is indicative of a
true evolution of  the metallicity in individual systems
(see also Prochaska \& Wolfe 2000; Savaglio et al. 1999).  
This should be  checked by measuring in
the same systems  abundances of species like carbon,  aluminium, silicon and
iron.

\begin{table}
\begin{tabular}{lllll}
\multicolumn{5}{l}{{\bf Table 3.} Column densities$^a$ in the 
$z_{\rm abs}$~=~2.974 system}\\ 
\hline
\multicolumn{1}{c}{Species}&
\multicolumn{1}{c}{log~$N$}&\multicolumn{1}{c}{$Z^b$}&
\multicolumn{1}{c}{$Z_{\odot}$}&
\multicolumn{1}{c}{[X/H]$^{b,c}$}\\
\multicolumn{1}{c}{}
&\multicolumn{1}{c}{(cm$^{-2}$)}&\multicolumn{1}{c}{}&
\multicolumn{1}{c}{}\\
\hline
H~I & 19.8--20.3 & & & \\
C~{\sc ii}  & 14.5 & $-$5.80 & $-$3.45 & $-$2.35 \\
O~{\sc i}   & 14.8: & $-$5.50 & $-$3.13 & $-$2.37 \\
Al~{\sc ii} & 12.5 & $-$7.78 & $-$5.52 & $-$2.26 \\
Si~{\sc ii} & 13.8 & $-$6.55 & $-$4.45 & $-$2.10 \\
Fe~{\sc ii} & 13.5 & $-$6.80 & $-$4.49 & $-$2.31 \\
Zn~{\sc ii} & $<$11.8 & $<-$8.50 & $-$7.35 & $<-$1.1 \\
Ni~{\sc ii} & $<$12.5 & $<-$7.80 & $-$5.75 & $<-$2.0 \\
\hline
\multicolumn{5}{l}{$^a$ logarithm of, in cm$^{-2}$; $^b$
log~$N$~(H~{\sc i})~=~20.3 is assumed; }\\
\multicolumn{5}{l}{$^c$ typical errors are $\pm$0.3~dex.}\\
\label{tab297}
\end{tabular}
\end{table}
\section{Conclusion}
The doublet ratio of several intervening Mg~{\sc ii}$\lambda\lambda$2796,2803  
systems along the line of sight to APM~08279+5255
is observed close to unity, indicating saturation of the lines, 
whereas the depth  of the lines is close to 0.5 in the normalized spectrum 
(see Fig.~\ref{cf}).  
This can be understood if the absorption profile is made of  
components with arbitrarily small Doppler parameters ($b$~$\sim$~1--1.5~km~s$^{-1}$).
This would imply however a surprisingly low temperature (1500--3000~K)
when the gas is 
expected to be heated by photo-ionization to temperatures larger
than 10$^{4}$~K (e.g. Petitjean et al. 1992).
%
%
A more likely explanation of these observations is that Mg~{\sc ii} galactic 
halos are composed of a collection of clouds each of them having dimensions 
less than $\sim$1~kpc. Individual clouds, regularly spread over the velocity 
profile by kinematics, cover only one of the two brighest image of the lensed 
quasar. The number density of clouds is 
not large enough for the absorption material to cover the two lines of sight 
at all velocity positions. The total covering factor of the halo however 
is close to one, consistent with observations of associated galaxies at
intermediate redshift. In  contrast,  the two  strong  Mg~{\sc ii}  systems 
at  $z_{\rm abs}$~=~1.06 and  1.18 have covering  factor equal to  one (the
lines   are  saturated  and   go  to   the  zero   level)  over   more  than
200~km~s$^{-1}$.  These latter systems are likely to arise due to absorption
through the central part of galaxies where the number of clouds is  so large 
that saturated absorption occurs along both lines of sight whatever the 
radius of the individual clouds might be.

Models by Mo \& Miralda-Escud\'e (1996) have shown that 
halos with small rotation velocity ($<$~100~km~s$^{-1}$) 
should contribute little to the 
total cross-section of Mg~{\sc ii} systems as they have
dimensions as small as 5~kpc. This conclusion is probably true
at $z$~$<$~1 (see also Churchill et al. 1996).
It is however interesting to note that most of the $z$~$>$~1 Mg~{\sc ii} systems 
studied here (see Fig.~\ref{cf}) are spread over much less than 
100~km~s$^{-1}$. Moreover,
they have equivalent widths $W_{\rm r}$~$\sim$~0.43, 0.37, 0.12,
0.99, 0.28 and 0.17~\AA~ at $z_{\rm  abs}$~=~1.211, 1.5497, 1.5523, 1.813, 
2.0418 and 2.0668 respectively. 
Therefore the systems we see have similar strengths as the systems which, at 
lower redshift, are associated with large halos of galaxies. 
In particular, they are generally stronger than the weak Mg~{\sc ii} systems 
studied by Churchill et al. (1999). 
This means that, contrary to what is seen at lower redshift, these systems 
could be associated with halos of low rotation velocity ($<$~100~km~s$^{-1}$)
and thus small radii. 

The two strong  Mg~{\sc ii}  systems at  $z_{\rm abs}$~=~1.06 and 1.18 are 
studied in detail.
Absorption from Ca~{\sc ii},  Mg~{\sc i}, Ti~{\sc ii}, Mn~{\sc ii} and
Fe~{\sc ii} have been observed in several damped Lyman-$\alpha$ systems over
a large range of redshift (Meyer et al. 1995, Lu et al. 1996, Vladilo et al.
1997, Proschaska \& Wolfe 1997,  Churchill et al. 2000).  This is, however,
the first time  that Na~{\sc i} is also detected at such redshift,  
thanks to the combination of  high S/N  ratio and  high spectral  resolution.  
This  additional strong
constraint shows that the gas in  these systems is cool and neutral. Indeed,
similar column densities are observed in our Galaxy for Ca~{\sc ii}, Mg~{\sc i},
Ti~{\sc ii}, Mn~{\sc ii} and Fe~{\sc  ii} in warm and cool gas clouds toward,
respectively, $\mu$Col and  23~Ori.  
Only the  Na~{\sc  i} column
density differs;  it is more than  an order of magnitude  larger through the
cool  cloud.   Doppler parameters  as  low  as $b$~$\sim$~1~km~s$^{-1}$  are
derived from Voigt-profile fitting  of isolated subcomponents.  We find that
the  H~{\sc  i} column  density  at  $z$~=~1 is  of  the  order of  $1\times
10^{21}$~cm$^{-2}$   to  $5\times   10^{21}$~cm$^{-2}$,   the  corresponding
metallicity is  in the range 1--0.3~$Z_{\odot}$, the dust-to-metal ratio
is about a third that in our Galaxy and the relative depletions
of iron, titanium, manganese and calcium are similar to those
in cool gas in the  disk of our Galaxy.  
The dust-to-metal ratio measured here is similar to what is derived in most 
of the damped Lyman-$\alpha$ systems (Vladilo 1998,  Savaglio et al.  1999).  
The presence of dust is supported by the reddening of the QSO spectrum over the 
$R$-band.
These   are  probably   amongst  the   most  metal   and   dust-rich  damped
Lyman-$\alpha$ systems  at $z$~$\sim$~1.  
The dust  depletion pattern is similar to  that observed in cool  gas in the
Galaxy. All this  is
consistent with  the finding by  Petitjean et al.   (1992) that although
most  of the damped Lyman-$\alpha$ systems arise in warm gas, the 
highest column densities are due to 
a collection of clumps that condense out of the 
warm phase due to thermal instability (see also Lane et al. 2000).

Another damped  Lyman-$\alpha$ system is seen at  $z_{\rm abs}$~=~2.974 with
19.8~$<$~log~$N$(H~{\sc i})~$<$~20.3.   As the Lyman-$\alpha$  line is black
over  about  15~\AA,  the  cloud  must  cover  the  three  QSO  images.  The
transverse dimension  of   the  absorber   is  therefore   larger  than
200~$h^{-1}_{75}$~pc.  Column densities of Al~{\sc ii}, Fe~{\sc ii}, Si~{\sc
ii},  C~{\sc ii}  and O~{\sc  i} indicate  abundances relative  to  solar of
$-$2.31, $-$2.26,  $-$2.10, $-$2.35 and  $-$2.37 for, respectively,  Fe, Al,
Si,  C and O  (for log~$N$(H~{\sc  i})~=~20.3). Metallicities  are therefore
less than 10$^{-1.5}$~$Z_{\odot}$ and, if any, the amount of dust in the cloud
is very small, as are any deviations from relative solar abundances. 
It seems likely that the difficulty to detect Zinc in
several damped Lyman-$\alpha$ systems at $z$~$>$~3 in previous surveys is 
indicative of a true cosmological evolution of  the metallicity in individual
systems (see also Prochaska \& Wolfe 2000, Savaglio et al. 1999). This should 
be  checked by measuring in the same systems  abundances of species like carbon,  
aluminium, silicon and iron.

\acknowledgements{We thank the team headed by
Sara L. Ellison to have made this beautiful data available for
general public use and particularly Sara L. Ellison for an access
to individual spectra. We gratefully acknowledge support from the Indo-French 
Centre for the Promotion of Advanced Research (Centre Franco-Indien 
pour la Promotion de la Recherche Avanc\'ee) under contract No. 1710-1.
PPJ thanks Elisabeth Flam for useful discussions and
Patrick Boiss\'e for a critical reading of the manuscript.
}


\begin{thebibliography}{}
\bibitem{}
Bergeron J., Boiss\'e P., 1991, A\&A 243, 344
\bibitem{}
Bohlin R.C., Savage B.D., Drake J.F., 1978, ApJ 224, 132
\bibitem{}
Bowen D.V., Blades J.C., Pettini M., 1995, ApJ 448, 634
\bibitem{}
Carswell R.F., Webb J.K., Baldwin J.A., Atwood B., 1987, ApJ 319, 709
\bibitem{}
Churchill C.W., 1999, astro-ph/9909426
\bibitem{}
Churchill C.W., Mellon R.R., Charlton J.C., et al., 2000, ApJ submitted
\bibitem{}
Churchill C.W., Rigby J.R., Charlton J.C., Vogt S.S., 1999, ApJS 120, 51
\bibitem{}
Churchill C.W., Steidel C.C., Vogt S.S., 1996, ApJ 471, 164
\bibitem{}
Diplas A., Savage B.D., 1994, ApJS 93, 211
\bibitem{}
Ellison S.L., Lewis G.F., Pettini M., et al.
 1999a, PASP 111, 946
\bibitem{}
Ellison S.L., Lewis G.F., Pettini M., et al. 
1999b, ApJ 520, 456
\bibitem{}
Elvis M., Wilkes B.J., McDowell J.C., et al., 1994, ApJS 95, 1
\bibitem{}
Ferland, G. J. 1996, "HAZY a Brief Introduction to Cloudy", Univ.
Kentucky, Dept. Physics \& Astron.. Internal rep.
\bibitem{}
Ferlet R., Vidal-Madjar A., Gry C., 1985, ApJ 298, 838
\bibitem{}
Fontana A., Ballester P., 1995, The Messenger 80, 37
\bibitem{}
Gredel R., van Dishoeck E.F., Black J.H., 1993, A\&A 269, 477
\bibitem{}
Haehnelt M.G., Steinmetz M., Rauch M., 1998, ApJ 495, 647
\bibitem{}
Hobbs L.M., 1978, ApJ 222, 491
\bibitem{}
Howk J.C., Savage B.D., Fabian D., 1999, ApJ 525, 253
\bibitem{}
Ibata R.A., Lewis G.F., Irwin M.J., Leh\'ar J., Totten E.J.,
1999, AJ 118, 1922
\bibitem{}
Irwin M.J., Ibata R.A., Lewis G.F., Totten E.J., 1998, ApJ 505, 529
\bibitem{}
Lane W.M., Briggs F.H., Smette A., 2000, ApJ 532, L146 
\bibitem{}
Ledoux C., Petitjean P., Bergeron J., Wampler E.J., Srianand R., 1998a, A\&A
337, 51
\bibitem{}
Ledoux C., Th\'edore B., Petitjean P., et al., 1998b, A\&A 339, L77
\bibitem{}
Lespine Y., Petitjean P., 1997, A\&A 317, 416
\bibitem{}
Lewis G.F., Chapman S.C., Ibata R.A., Irwin M.J., Totten E.J., 1998,
ApJ 505, L1
\bibitem{}
Lopez S., Reimers D., Rauch M., Sargent W.L.W., Smette A., 1999,
ApJ 513, 598
\bibitem{}
Lu L., Sargent W.L.W., Barlow T.A., Churchill C.W., Vogt S.S., 1996, ApJS
107, 475
\bibitem {}
Meyer D.M., Lanzetta K.M., Wolfe A.M., 1995, ApJ 451, L13
\bibitem {}
Meyer D.M., Lauroesch J.T., 1999, ApJ 520, L103
\bibitem {}
Mo H.J., Miralda-Escud\'e J., 1996, ApJ 469, 589
\bibitem {}
Monier E.M., Turnshek D.A., Lupie O.L., 1998, ApJ 496, 177
\bibitem{}
Pei Y.C., Fall S.M., Bechtold J., 1991, ApJ 378, 6
\bibitem{}
P\'equignot D., Aldrovandi S.M.V., 1986, A\&A 161, 169
\bibitem{}
Petitjean P., Bergeron J., 1990, A\&A 231, 309
\bibitem{}
Petitjean P., Bergeron J., Puget J.L., 1992, A\&A 265, 375
\bibitem{}
Petitjean P., Rauch M., Carswell R.F., 1994, A\&A 291, 29
\bibitem{}
Pettini M., Ellison S.L., Steidel C.C., Bowen D.V., 1999, ApJ 510, 576
\bibitem{}
Pettini M., Smith L.J., King D.L., Hunstead R.W., 1997, ApJ 486, 665
\bibitem{}
Prochaska J.X., Wolfe A.M., 1997, ApJ 474, 140 
\bibitem{}
Prochaska J.X., Wolfe A.M., 1998, ApJ 507, 113 
\bibitem{}
Prochaska J.X., Wolfe A.M., 2000, astro-ph/0002513
\bibitem{}
Rauch M., Sargent W.L.W., Barlow T.A., 1999, ApJ 515, 500
\bibitem{}
Savage B.D., Sembach K.R., 1996, ARAA 34, 279
\bibitem{}
Savaglio S., Panagia N., Stiavelli M., 1999, astro-ph/9912112
\bibitem{}
Seaton M.J., 1979, MNRAS 187, 
\bibitem{}
Sembach K.R., Danks A.C., 1994, A\&A 289, 539
\bibitem{}
Sembach K.R., Danks A.C., Savage B.D., 1993, A\&AS 100, 107
\bibitem{}
Smette A., Robertson J.G., Shaver P.A., et al., 1995, A\&AS 113, 199
\bibitem{}
Srianand R., Khare P.,1994, ApJ 428, 82
\bibitem{}
Srianand R., Kembhavi A., 1997, ApJ 478, 70
\bibitem{}
Srianand R., Petitjean P., 2000, A\&A in press; astro-ph/0003301
\bibitem{}
Srianand R., Shankaranarayanan S., 1999, ApJ 518, 672
\bibitem{}
Steidel C.C., 1993, Shull J.M. \& Thronson H.A., The Environment and 
Evolution of Galaxies, Proc. of the 3rd Teton Astronomy Conference, Dordrecht,
Kluwer, p. 263
\bibitem{}
Stockes G.M., 1978, ApJS 36, 115 
\bibitem {}
Th\'evenin F., 1998, Catalogue III/193, Bull. CDS, Vol. 49
\bibitem {}
Vidal-Madjar A., Andreani P., Cristiani S., et al., 1987, A\&A 177, L17
\bibitem {}
Vladilo G., 1998, ApJ 493, 583
\bibitem {}
Vladilo G., Molaro P., Monai S., et al., 1993, A\&A 274, 37
\bibitem {}
Vladilo G., Centuri\'on M., Falomo R., Molaro P., 1997, A\&A 327, 47
\bibitem {}
Welty D.E., Hobbs L.M., Kulkarny V.P., 1994, ApJ 436, 152
\bibitem {}
Welty D.E., Hobbs L.M., Lauroesch J.T., Morton D.C., Spitzer L., York D.G.,
1999, ApJS 124, 465
\bibitem {}
Yun M.S., Ho P.T.P., Lo K.Y., 1994, Nature 372, 530
\bibitem {}
Zheng W., Kriss G.A., Telfer R.C., et al., 1997, ApJ 475, 469
\end{thebibliography}
\end{document}